\documentclass[twocolumn]{aastex631}
\shorttitle{Mixing diversity and helium core masses}
\shortauthors{Pedersen}

\usepackage{amsmath}

\begin{document}

\title{\large On the diversity of mixing and helium core masses of B-type dwarfs from gravity-mode asteroseismology}

\email{mgpedersen@kitp.ucsb.edu}

\author[0000-0002-7950-0061]{May G. Pedersen}
\affiliation{Kavli Institute for Theoretical Physics, Kohn Hall, University of California, 
Santa Barbara, CA 93106, USA}



\begin{abstract}

\noindent The chemical evolution of the Galaxy is largely guided by the yields from massive stars. Their evolution is heavily influenced by their internal mixing, allowing the stars to live longer and yield a more massive helium core at the end of their main-sequence evolution. Asteroseismology is a powerful tool for studying stellar interiors by providing direct probes of the interior physics of the oscillating stars. This work revisits the recently derived internal mixing profiles of 26 slowly pulsating B stars observed by the \emph{Kepler} space telescope, in order to investigate how well the mixing profiles can in fact be distinguished from one another as well as provide predictions for the expected helium core masses obtained at the end of the main-sequence evolution. We find that for five of these stars the mixing profile is derived unambiguously, while the remaining stars have at least one other mixing profile which explains the oscillations equally well. 
Convective penetration is preferred over exponential diffusive overshoot for $\approx$55\% of the stars, while stratified mixing is preferred in the envelope ($\approx$39\%).
We estimate the expected helium core masses obtained at the end of the main-sequence evolution and find them to be highly influenced by the estimated amount of mixing occurring in the envelopes of the stars. 
\end{abstract}



\section{Introduction} \label{sec:intro}

\noindent The calculation of stellar chemical yields, i.e. the fraction of the initial stellar mass returned to the interstellar medium in the form of newly produced elements during the entire stellar lifetime \citep{Tinsley1980}, are crucial for predicting the chemical evolution of the Galaxy, and defines the chemical composition of newly formed stars. These chemical yield calculations depend on the mass loss from stellar winds and the contribution from supernovae explosions \citep[e.g.][]{Hirschi2005}. 

For stars more massive than 8-10\,M$_\odot$, the formation of a black hole through a core collapse can occur with or without a supernova explosion \citep{Woosley2002}. This latter case is known as a failed core collapse supernova (CCSN). \cite{Adams2017a} estimated that about 14\% of all red supergiants will experience such a failed CCSN, while other studies put an upper limit at 50\% \citep{Lien2010,Horiuchi2011,Horiuchi2014}. In a failed CCSN of a red supergiant, the remnant left behind has the same mass as the pre-collapse helium core \citep{Nadezhin1980,LovegroveWoosley2013}. These pre-collapse helium core masses $m_{\rm He}$ are critically dependent on the internal chemical mixing history of the stars, especially on the main-sequence where the most time is spent.

For stars born with convective cores on the main-sequence, additional mixing can occur in both the convective core boundary layer, referred to as convective boundary mixing (CBM), as well as in the radiative envelope.
The effect of the CBM is to bring additional hydrogen fuel to the convective core where it is very efficiently mixed, allowing the stars to live longer and have a larger final $m_{\rm He}$ at the end of the main-sequence, and has been studied for several decades \citep[see e.g.][for a few examples]{Roxburgh1965,Saslaw1965,Maeder1976,Maeder1981,Bressan1981,Stothers1985,Bertelli1985,Chin1991,Umezu1995,Browning2004,Rosenfield2017,Johnston2019b,Kaiser2020}. Similarly, envelope mixing processes from rotation \citep{Maeder2009,Ekstrom2012,Georgy2013} and internal gravity waves \citep{Rogers2017} can bring  new hydrogen to the stellar core and transport helium and excess nitrogen produced in the CNO cycle to the surface. The study of the surface abundances of helium, carbon, nitrogen, and oxygen and their relative ratios can then be used as an observational diagnostic of the internal envelope mixing of stars \citep[e.g.][]{Hunter2009,Fraser2010,Przybilla2010,Brott2011,Bouret2013,Maeder2014,Martins2015,Grin2017,Bouret2021}.

Asteroseismology provides a powerful tool for studying the interiors of stars, as stellar oscillations penetrate deep inside the stars and carry information about the conditions within. Two types of oscillators are found on the upper main-sequence: the $\beta$ Cephei ($\beta$~Cep) and the slowly pulsating B (SPB) stars. 
With its four years of high-precision, high-cadence continuous observations, the nominal \emph{Kepler} mission \citep{Borucki2010} provides the best photometric data for studying stellar oscillations available to date. Because of its choice of field-of-view, the highest mass main-sequence stars observed by \emph{Kepler} are the SPB stars.

The SPB stars have masses in the range 3-10\,M$_\odot$ and oscillate in gravity (g) modes, where gravity acts as the dominant restoring force for the oscillations \citep{Aerts2010}. Until recently, the extent of the CBM region has been estimated for 17 SPB stars \citep{Degroote2010,Szewczuk2015,Moravveji2015,Moravveji2016,Szewczuk2018,Johnston2019,Mozdzierski2019,Wu2019,Fedurco2020,Wu2020,Michielsen2021}, and the amount of envelope mixing determined for four of them \citep{Moravveji2015,Moravveji2016,Wu2019,Wu2020}.
For only two of the stars were different shapes of the CMB mixing tested against the observations \citep{Moravveji2015,Moravveji2016,Michielsen2021}, while \cite{Michielsen2021} tested the temperature gradient within the CMB region for one SPB star. In all three cases, the shape of the envelope mixing profile was held fixed. In their latest study, \cite{Pedersen2021} performed asteroseismic modeling of 26 SPB stars and tested which out of eight different shapes of the internal mixing profile did the best at reproducing the observed oscillations. This is both the highest number of SPB stars and highest number of profiles that have been modeled asteroseismically in a single homogeneous study to date, and it showed that the mixing among the 26~stars is very diverse. 

While the work by \cite{Pedersen2021} took an important step towards unraveling the internal mixing of stars with convective cores by determining which of the eight considered profiles provided the best matching theoretical predictions to the observed g-mode characteristics of the 26 SPB stars, they did not investigate to what extent these internal mixing profiles could be distinguished from the rest. Here we remedy this situation by investigating if these 26 derived internal mixing profiles were determined unambiguously. We further expand upon the study of \cite{Pedersen2021} by estimating the final $m_{\rm He}$ values obtained at the end of main-sequence evolution for these 26~SPB stars, and studying to what extent these core masses are influenced by the internal mixing of the stars.

We provide an overview of the mixing profiles considered by \cite{Pedersen2021} in Sect.~\ref{Sec:Dmix_overview}, and discuss the ability of the SPB stars to distinguish between these eight profiles in Sect.~\ref{sec:mixing}. Based on the best model estimates from \cite{Pedersen2021} and their computed eight grids of stellar models, we predict the final helium core masses obtained at the end of the main-sequence evolution in Sect.~\ref{sec:Hecore}
and conclude in Sect.~\ref{sec:Conclusions}.


\section{Internal mixing profiles}\label{Sec:Dmix_overview}

\noindent For the asteroseismic modelling of 26 SPB stars, \cite{Pedersen2021} computed eight different grids of stellar models each assuming their own shape of the internal mixing profile. 
The stellar models $\mathcal{M} \left(\boldsymbol{\psi}, \boldsymbol{\theta} \right)$ were computed using the stellar structure and evolution code \texttt{MESA} \citep{Paxton2011,Paxton2013,Paxton2015,Paxton2018,Paxton2019} and their theoretical oscillations derived using the stellar oscillation code \texttt{Gyre} \citep{Townsend2013,Townsend2018,Goldstein2020}. The free parameters are represented by the vector $\boldsymbol{\theta}$, while $\boldsymbol{\psi}$ denotes the fixed input physics. The choice of the shape of the internal mixing profile corresponds to one choice of $\boldsymbol{\psi}$, and for the sake of simplicity $\boldsymbol{\psi}_1, \dots, \boldsymbol{\psi}_8$ are used to represent the eight different grids. We refer to \cite{Pedersen2021} for the specifics behind the other fixed input physics in the \texttt{MESA} and \texttt{Gyre} setups. 

The internal mixing is treated as a diffusive process and represented by the diffusive mixing coefficient, $D_{\rm mix} (r)$, which defines the efficiency of the mixing as a function of the radius coordinate $r$. The diffusive mixing coefficient is split into three different components corresponding to the contribution to the mixing from the convective regions, $D_{\rm conv} (r)$, the envelope $D_{\rm env} (r)$, and the boundary layer between the convective core and the envelope, $D_{\rm cbl} (r)$.

\cite{Pedersen2021} considered two different shapes for $D_{\rm cbl} \left(r \right)$. The first CBM profile uses the exponential diffusive overshoot prescription \citep{Freytag1996,Herwig2000}, which assumes a radiative temperature gradient $\nabla_{\rm rad}$ in the CBM region and the mixing efficiency exponentially decreases with the distance from the convective core boundary. How rapidly the mixing efficiency decreases is set by the free parameter $f_{\rm ov}$, which thereby sets the extent of the CBM region. For the second $D_{\rm cbl} (r)$, \cite{Pedersen2021} used convective penetration \citep{Zahn1991}, which adopts the adiabatic temperature gradient $\nabla_{\rm ad}$ in the CBM region and assumes that the mixing is constant over a distance $\alpha_{\rm pen} H_{\rm p}$ from the convective core boundary. Here $H_{\rm p}$ is the pressure scale height and $\alpha_{\rm pen}$ is the free parameter.

Four different shapes of $D_{\rm env} (r)$  were considered by \cite{Pedersen2021}, each with the same free parameter $D_{\rm env, 0}$, which sets the diffusive mixing coefficient at the change from $D_{\rm cbl} \left(r \right)$ to $D_{\rm env} (r)$. For the first envelope mixing profile, the mixing is set to be constant throughout the envelope, $D_{\rm env} (r) = D_{\rm env, 0}$. This is what was previously used to model two SPB stars by \cite{Moravveji2015,Moravveji2016}. The second profile follows the prediction of diffusive mixing from internal gravity waves \citep[IGWs,][]{Rogers2017} for which $D_{\rm env} (r) = D_{\rm env, 0} \left(\rho/\rho_0 \right)^{-1/2}$, where $\rho$ is the density and $\rho_0$ is the value of the density at the switch from $D_{\rm cbl} \left(r \right)$ to $D_{\rm env} (r)$. The last two envelope mixing profiles assume rotational mixing from different mechanisms. The first uses mixing due to vertical shear instabilities \citep{Mathis2004}, while the second combines mixing arising from a combination of meridional circulation and large horizontal and vertical shear \citep{Georgy2013}. The two profiles were taken from \cite{Georgy2013} and scaled to $D_{\rm env, 0}$ at the position of $\rho_0$.

Combining the two $D_{\rm cbl} (r)$ and four $D_{\rm env} (r)$ profiles results in a total of eight different internal mixing profiles. These profiles are summarized in Table~\ref{Tab:Dmix} in this work as well as in Fig.~3 of \cite{Pedersen2021}.

\begin{deluxetable*}{cccc}
\tablecaption{Summary of the eight grids of stellar models computed with \texttt{MESA} and resulting from the combinations of two different mixing profiles in the convective boundary layer $D_{\rm cbl} (r)$ and four different envelope mixing profiles $D_{\rm env} (r)$.\label{Tab:Dmix}}
\tablewidth{700pt}
\tabletypesize{\small}
\tablehead{
\colhead{Grid} & \colhead{$D_{\rm cbl} (r)$} & \colhead{$D_{\rm env} (r)$} & \colhead{Free mixing parameters} 
} 
\startdata
 	$\psi_1$	&   Exponential overshoot 	&  	Constant		& $f_{\rm ov}$; $D_{\rm env, 0}$\\[0.5ex]  
	$\psi_2$	&   Exponential overshoot 	&  	IGWs		& $f_{\rm ov}$; $D_{\rm env, 0}$\\[0.5ex]  
	$\psi_3$	&   Exponential overshoot 	&  	Vertical shear		& $f_{\rm ov}$; $D_{\rm env, 0}$\\[0.5ex]  
	$\psi_4$	&   Exponential overshoot 	&  	Meridional circulation + vertical shear		& $f_{\rm ov}$; $D_{\rm env, 0}$\\[0.5ex]  
	$\psi_5$	&   Convective penetration	&  	Constant		& $\alpha_{\rm pen}$; $D_{\rm env, 0}$\\[0.5ex]  
	$\psi_6$	&   Convective penetration 	&  	IGWs		& $\alpha_{\rm pen}$; $D_{\rm env, 0}$\\[0.5ex]  
	$\psi_7$	&   Convective penetration 	&  	Vertical shear		& $\alpha_{\rm pen}$; $D_{\rm env, 0}$\\[0.5ex]  
	$\psi_8$	&   Convective penetration 	&  	Meridional circulation + vertical shear		& $\alpha_{\rm pen}$; $D_{\rm env, 0}$\\[0.5ex]  
 \enddata
\end{deluxetable*}


\section{Differentiation capability between different internal mixing profiles}\label{sec:mixing}

\noindent The asteroseismic modeling carried out by \cite{Pedersen2021} compared observed g-mode period spacing patterns to theoretical patterns predicted from stellar models of varying masses $M$, main-sequence ages\footnote{The current central hydrogen mass fraction $X_{\rm c}$ with respect to the initial one $X_{\rm ini}$ is used as a proxy for the age of the star.} $X_{\rm c}/ X_{\rm ini}$, metallicities $Z$, and mixing parameters ($f_{\rm ov}$ or $\alpha_{\rm pen}$, $D_{\rm env,0}$), with the effect of the rotational frequency $f_{\rm rot}$ of the star on the patterns taken into account in the \texttt{Gyre} computations where uniform rotation was assumed. While differential rotation can cause slight changes to the morphology of the period spacing pattern, \cite{VanReeth2018} demonstrated that multiple period spacing patterns are required to detect differential rotation unless the star is highly differentially rotating\footnote{More than a 90\% difference between core and surface rotation rate.}. Furthermore, Fig. 4 by \cite{Aerts2019b} shows that the majority of stars with convective cores on the main-sequence are rigidly rotating. Therefore, we consider the assumption of uniform rotation to be an appropriate approach for the asteroseismic modeling of the 26 SPB stars.

The period spacing patterns are built from the periods of stellar g-mode  oscillations that have the same degree $\ell$ and azimuthal order $m$ and are consecutive in radial order $n$. Calculating the period differences $\Delta P$ between two such oscillations that are consecutive in $n$ (i.e. have $\Delta n = 1$) and plotting them as a function of the period of the oscillations $P$, is what we call a period spacing pattern. The morphology of these patterns change depending on the mass, age, mixing, temperature gradient, and rotation of the star \citep{Miglio2008,Bouabid2013,VanReeth2015,Pedersen2018,Michielsen2019}, and thereby allows us to constrain these stellar properties by modeling the patterns. One period spacing pattern was modeled for each of the 26~SPB stars, and they were all built from dipole ($\ell = 1$) modes, see Fig.~\ref{fig:pattern} in Appendix~\ref{Sec:KIC4930889} for an example of an observed period spacing pattern.

\cite{Pedersen2021} relied on statistical models to approximate the theoretical period spacing values as a function of the six varied stellar parameters $\boldsymbol{\theta} = \left(M, Z, X_{\rm c}/X_{\rm ini}, f_{\rm ov}\ {\rm or}\ \alpha_{\rm pen}, D_{\rm env, 0}, \Omega_{\rm rot}/\Omega_{\rm crit} \right)$\footnote{Here $\Omega_{\rm rot} = 2\pi f_{\rm rot}$ is the angular rotation frequency, and $\Omega_{\rm crit}$ its corresponding critical value assuming the Roche formalism, $\Omega_{\rm crit} =  \sqrt{8GM / 27  R^3}$. In this case, $R$ specifically refers to the equatorial radius.}. By doing so they were able to increase the resolution of the grid of theoretical models within the error range of the derived spectroscopic stellar parameters (effective temperature $T_{\rm eff}$, surface gravity $\log g$, luminosity $L$, and metallicity [M/H]) without having to perform any additional \texttt{MESA} or \texttt{Gyre} computations. We refer to \cite{Pedersen2021} for the specific details.

The statistical models are represented as

\begin{equation}
\Delta P_{ij} = \boldsymbol{x}_{ij}^\top \boldsymbol{\beta}_i,
	\label{Eq:stat_mod}
\end{equation}

\noindent where $\Delta P_{ij}$ is the $i$th period spacing value\footnote{The $i$ in this case corresponds to a given combination of $\ell$, $m$, and $n$ such that $\Delta P_i = P_{\ell,m,n} - P_{\ell, m, n-1}$.} at the $j$th grid point, $\boldsymbol{x}_{ij}$ is a list of fractional polynomials built from the components of $\boldsymbol{\theta}$, and $\boldsymbol{\beta}_i$ are the regression coefficients. 

Once the refined grid of models has been calculated using Eq.~(\ref{Eq:stat_mod}), these theoretical period spacing patterns are matched to the observed ones. The Mahalanobis distance (MD) at each grid point is then calculated, and the best matching model finally selected based on the Akaike's information criterion corrected for small number statistics, AICc \citep[see][]{Pedersen2021}. To derive the errors on the estimate $\boldsymbol{\theta}$ values, a Monte-Carlo approach was taken in which the regression coefficients $\boldsymbol{\beta}_j$ were perturbed based on their errors 100 times. For each of these 100 iterations the theoretical period spacing patterns were recalculated for all grid points and the best matching model redetermined. Here we increase the number of iterations from 100 to 1000, and redo the matching of the theoretical to the observed period spacing pattern at each iteration instead of assuming that the selected range in radial orders $n$ for each grid point $j$ remains the same between the iterations. An example of what the AICc distributions and theoretical period spacing patterns resulting from these 1000 iterations look like is shown in Fig.~\ref{fig:example_dist} in Appendix~\ref{Sec:KIC4930889}.

The 1000 perturbations of Eq.~(\ref{Eq:stat_mod}) likewise results in updated errors on the estimated parameters. We provide these in Table~\ref{Tab:theta_parameters} and \ref{Tab:theta_parameters_extra} in Appendix~\ref{Sec:theta_estimates}, and discuss how they are derived in Appendix~\ref{Sec:error_pert}.


\subsection{Comparison of AICc distributions}

\begin{figure*}
\begin{center}
\includegraphics[width=0.85\linewidth]{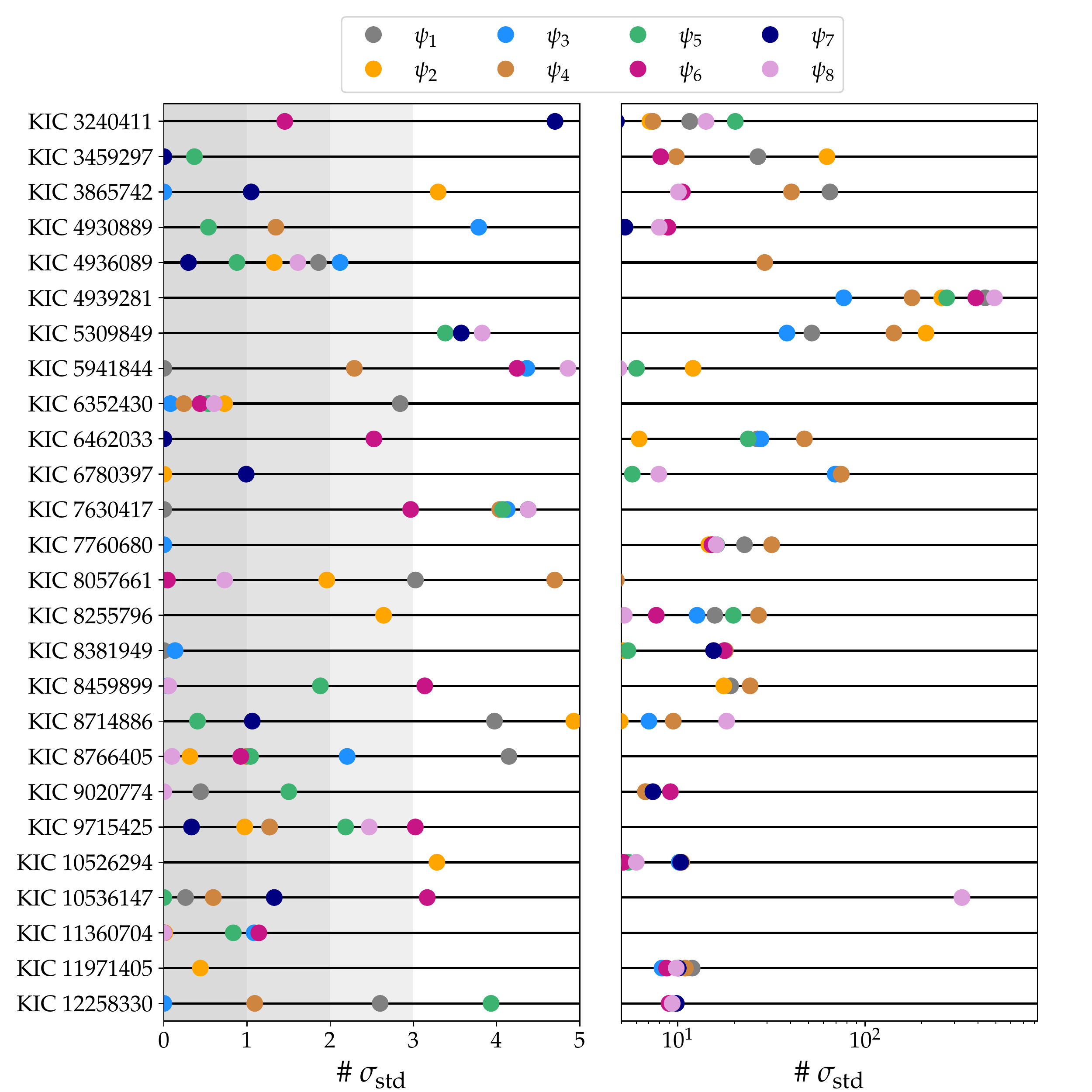}
\caption{Separation measured in the number of $\sigma_{\rm std}$ between the AICc distribution of the best matching model grid and the remaining seven grids of stellar models as calculated using Eq.~(\ref{Eq:sigma_std}). The left panel is a zoom in on the region where the separations are $\leq 5\,\sigma_{\rm std}^*$, while separations $\geq 5\,\sigma_{\rm std}^*$ are shown in the right hand panel on a logarithmic scale. The separations are plotted independently for each SPB star and for each grid as indicated by the colors of the data points.}\label{fig:stds_val}
\end{center}
\end{figure*}

\noindent The separation between the eight AICc distributions resulting from the 1000 iterations discussed above and representing the eight different choices of $D_{\rm mix} (r)$ indicates how well the  best solutions from each grid can be distinguished from the rest. To quantify this, we calculate the median AICc value for the AICc distribution for the overall best matching model grid and its 86th percentile, which corresponds to the $1\sigma$ standard deviation in the case of a Gaussian distribution, $\sigma_{\rm std}$. For simplicity, we will refer to this 86th percentile for the best model as $\sigma_{\rm std}^{*}$, which is effectively the 86th percentile of the AICc distribution subtracted by its median value. Similarly, we calculate the median and 86th percentile values of the remaining seven grids of stellar models and refer to this 86th percentile as $\sigma_{\rm std}^q$, where $q$ takes values between 2nd and 8th, corresponding to the second and eighth best matching $D_{\rm mix} (r)$.  We then calculate the separation between the best matching model distribution and distributions of the remaining seven grids as

\begin{equation}
\#\ \sigma_{\rm std} = \frac{{\rm Med} \left({\rm AICc} \right)^q - \sigma_{\rm std}^q - {\rm Med} \left({\rm AICc} \right)^*}{\sigma_{\rm std}^{*}},
	\label{Eq:sigma_std}
\end{equation}

\noindent where ${\rm Med} \left({\rm AICc} \right)^*$ and ${\rm Med} \left({\rm AICc} \right)^q$ corresponds to the median of the AICc distributions for the best matching model and the other $q$th model grid. Hence we measure the separation by the number of $\sigma_{\rm std}^{*}$ 
that the median of the best matching model distribution is separated from the individual 86th percentiles of the other seven AICc distributions. These separations are illustrated in Fig.~\ref{fig:stds_val} for each of the 26 SPB stars modeled by \cite{Pedersen2021}.

Figure~\ref{fig:stds_val} shows that our ability to differentiate between the best matching model and the remaining seven model grids is heavily star dependent. Two extreme cases are KIC~6352430 and KIC~4939281. For KIC~6352430 all of the other seven AICc distributions are separated from the one of the best matching $D_{\rm mix} (r)$ by less than $3\,\sigma_{\rm std}^*$, and six of the grids are separated from the best model by less than $1\,\sigma_{\rm std}^*$, meaning that their 86th percentile regions are overlapping with the one estimated for the best matching model (see the fifth row in Fig.~\ref{fig:example_dist} in Appendix~\ref{Sec:KIC4930889} for an example of this). We therefore conclude that we cannot differentiate between the eight model grids for KIC~6352430. In comparison, the AICc distributions of all of the other seven grids lie more than $3\,\sigma_{\rm std}^*$ away from the one of the best matching model for KIC~4939281, and we can say for certain that $\boldsymbol{\psi}_7$ does the best at reproducing the observed period spacing pattern. 

\begin{figure*}
\begin{center}
\includegraphics[width=0.9\linewidth]{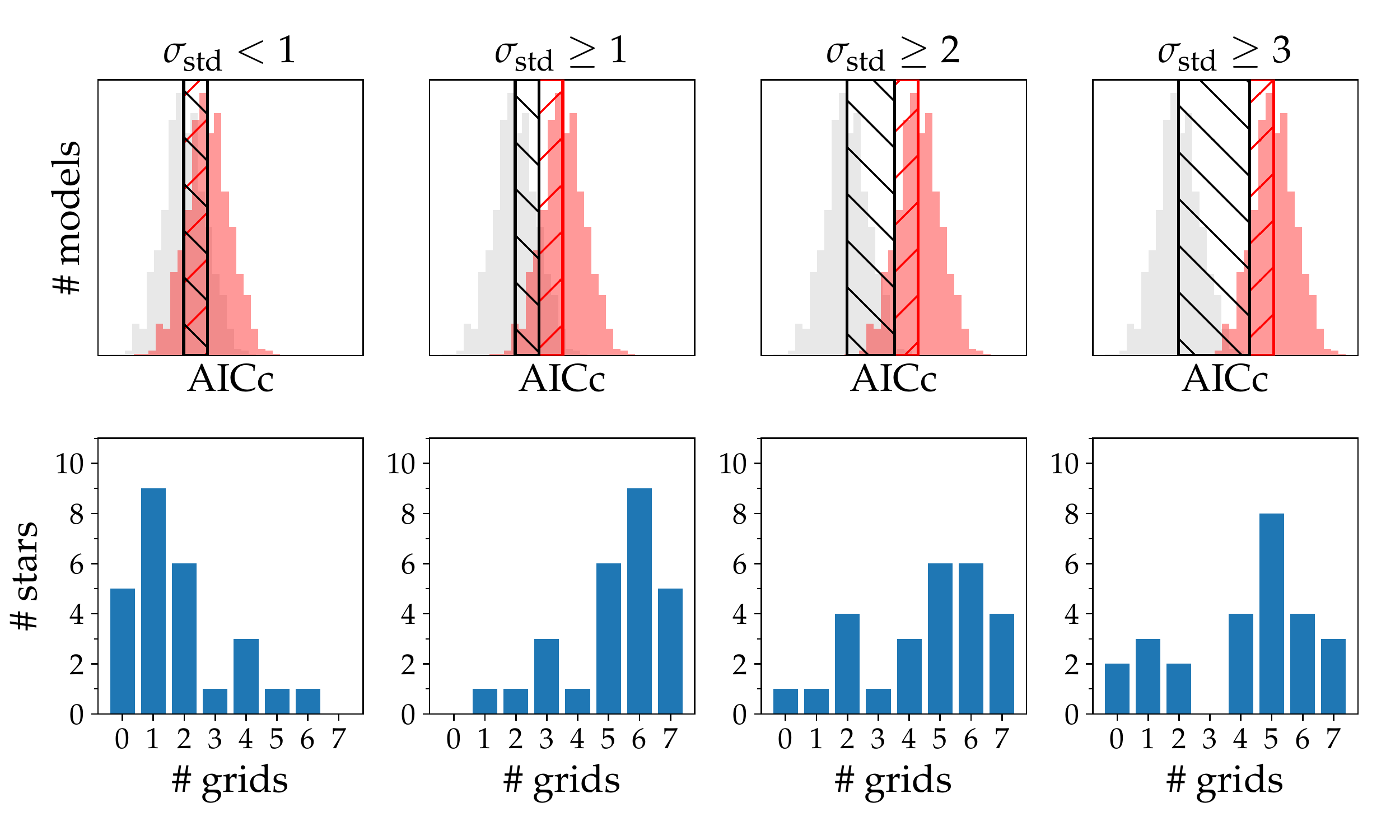}
\caption{Illustration of four different AICc distribution separation regimes (top panels) and their corresponding histograms (bottom panels). The histograms count the number of stars for which a certain number of grids are separated by from the best AICc distribution (indicated in light grey in the top panels) by less than $1\,\sigma_{\rm std}^*$ or more than  $1\,\sigma_{\rm std}^*$,  $2\,\sigma_{\rm std}^*$, or  $3\,\sigma_{\rm std}^*$ indicated by the black hatched region in the top panels. The AICc distribution in red corresponds to the distribution for one of the other seven model grids and its $1\,\sigma_{\rm std}^q$ region shown by the red hatching.}\label{fig:grid_stds_hist}
\end{center}
\end{figure*}

The results in Fig.~\ref{fig:stds_val} are summarized in the four histograms in the bottom panels of Fig.~\ref{fig:grid_stds_hist} as well as in Table~\ref{Tab:stds} in Appendix~\ref{Sec:Grid_stds}. The four different histograms consider four different separation regimes as illustrated in top row of panels in Fig.~\ref{fig:grid_stds_hist}. Here the AICc distribution in light grey corresponds to the one of the best matching model, while the second distribution in red shows the AICc distribution of one of the seven other considered model grids. The red hatched region always corresponds to the $\sigma_{\rm std}^q$ region of the red AICc distribution, while the black hatched region marks the $1\,\sigma_{\rm std}^{*}$, $1\,\sigma_{\rm std}^{*}$, $2\,\sigma_{\rm std}^{*}$, and $3\,\sigma_{\rm std}^{*}$ regions going from the left to the rightmost panel, respectively.

In the top left panel, the two hatched regions overlap, corresponding to the case where the separation between the distributions are less than $1\,\sigma_{\rm std}^*$. The histogram on the bottom panel counts the number of stars for which none (\# grids = 0) to all (\# grids = 7) of the other seven model grids are separated from the AICc distribution of the best matching model by less than $1\,\sigma_{\rm std}^*$. 
It shows, e.g., that none of the other seven model grids are separated by less than $1\,\sigma_{\rm std}$ from the best model distribution for five of the stars, and that no star has a separation $1 < \sigma_{\rm std}^*$ for all of the other seven grids simultaneously. 
The top right panel shows the case where the two distributions are separated by at least $3\,\sigma_{\rm std}^*$, with the corresponding histogram shown in the bottom panel.

\subsection{Constraints on mixing}

\noindent The results presented in Fig.~\ref{fig:stds_val} show that three of the SPB stars have all of the other seven grids separated from the best model AICc distribution by at least  $3\,\sigma_{\rm std}^*$. For these three stars we can therefore conclude that the estimated $D_{\rm mix} (r)$ from the best model grid provided by \cite{Pedersen2021} is indeed the one that best reproduces the observed period spacing pattern. For the remaining stars, there will always be some overlap between the AICc distribution of the best matching model and one to six of the other model grids. Fifteen of the stars have at least one or two grids which overlap with the best model distribution within $1\,\sigma_{\rm std}^*$, while 19 stars have the majority of the other grids separated from the median of best model distribution by at least  $3\,\sigma_{\rm std}^*$. 

\begin{figure}
\begin{center}
\includegraphics[width=0.9\linewidth]{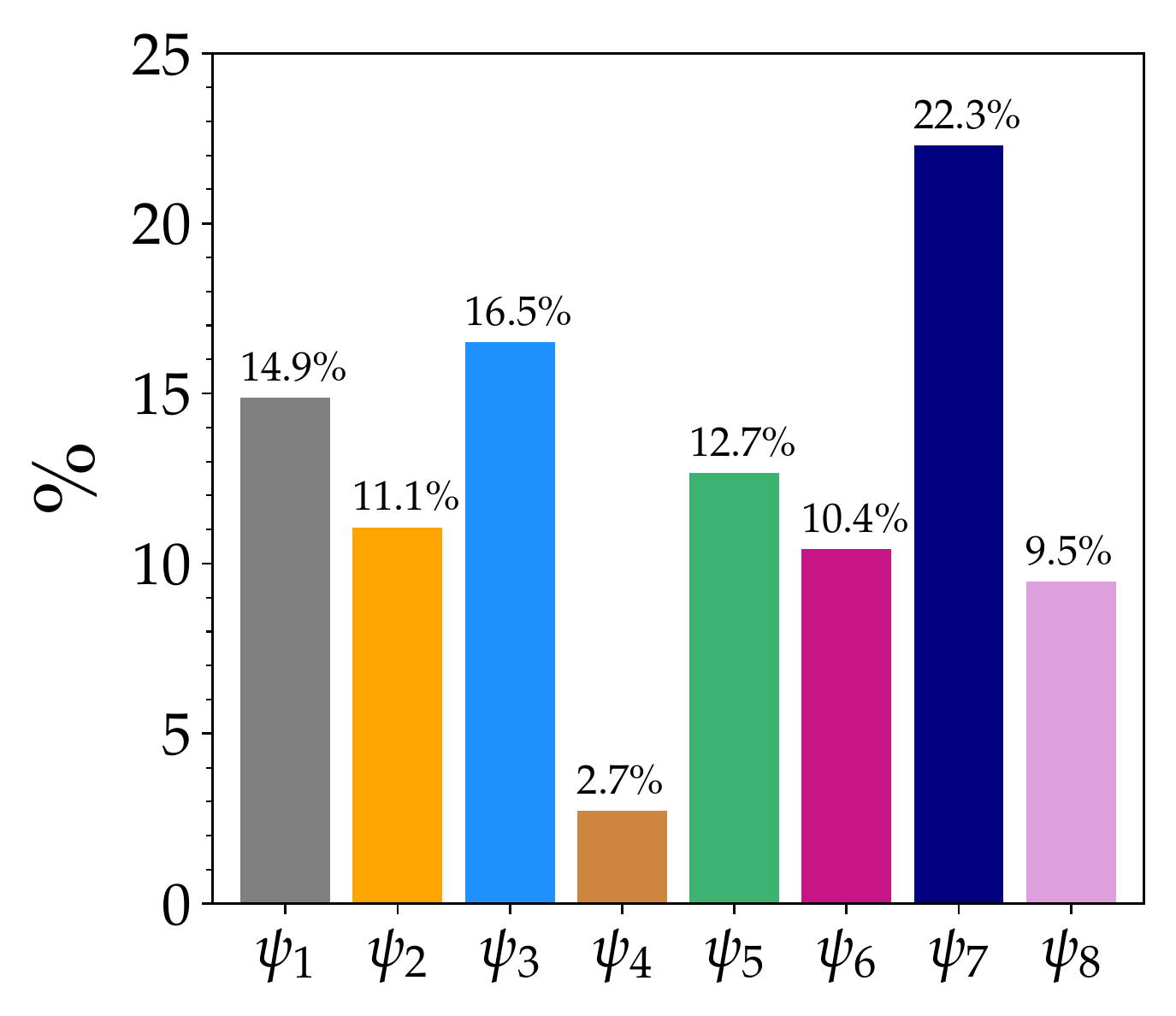}
\caption{Updated version of Fig.~4 from \cite{Pedersen2021}, with the number of stars in percentages preferring a given mixing profile now also accounting for the number of grids within $1\sigma_{\rm std}^*$ of the best matching one.}\label{fig:grid_hist}
\end{center}
\end{figure}

By counting the number of stars preferring a given mixing profile, \cite{Pedersen2021} found that the SPB stars generally prefer CBM from convective penetration and a stratified envelope mixing profile. Their calculations, however, did not account for how well the AICc distributions were separated from the one of the best matching model. Therefore we repeat the analysis here and remake the histogram in Fig.~4 of \cite{Pedersen2021}, now including per star  all of the models which match the observations equally well within $1\sigma_{\rm std}^*$, and dividing their contribution to the histogram by the total number of grids within $1\sigma_{\rm std}^*$ for the given star. The result is shown in Fig.~\ref{fig:grid_hist}. We find that 55\% of the stars prefer CBM from convective penetration over the exponential overshoot mixing profile, which is lower than the 65\% found by \cite{Pedersen2021}. For the envelope mixing profile, 39\% of the stars prefer the mixing profile derived from vertical shear instabilities, 28\% prefers a constant profile, 21\% prefers the mixing profile from IGWs, while 12\% of the stars prefer the envelope mixing profile from meridional circulation combined with large horizontal and vertical shear. In comparison to \cite{Pedersen2021}, the number of stars preferring the constant envelope mixing profile have increased by 10\% while the percentages for the other three $D_{\rm env} (r)$ profiles have decreased.

The asteroseismic modeling carried out by \cite{Pedersen2021} relied on the use of the statistical models in Eq.~(\ref{Eq:stat_mod}) to compare theoretical period spacing patterns to the observed patterns of the 26~SPB stars. These statistical models provide an estimate of the true theoretical period spacing patterns that one would obtain, if the theoretical oscillations were computed with \texttt{Gyre} from an underlying \texttt{MESA} model. It is to be expected that not all of the details from the real theoretical period spacing patterns can be predicted by the statistical models. Better distinction between the eight model grids can be made if a refined grid of \texttt{MESA} and \texttt{Gyre} models is computed for each $\boldsymbol{\psi}_1, \dots, \boldsymbol{\psi}_8$ mixing profiles and for each star, but this is a time consuming edeavor. Table~\ref{Tab:stds} therefore indicates for which stars such a study would have the highest impact, and where the priority should be given to the stars with the largest amount of overlap between the AICc distributions of the best matching model and the other seven grids. 

Finally, the overlap of the AICc distributions means that we have an additional error on the estimated parameters arising from the fact that we cannot distinguish between the $D_{\rm mix} (r)$ grids that are separated by less than $1\sigma_{\rm std}$. We therefore provide updated parameter estimates and errors in Tables~\ref{Tab:theta_parameters_averages} and \ref{Tab:theta_parameters_extra_averages} in Appendix~\ref{Sec:theta_estimates} and discuss how they are derived in Appendix~\ref{Sec:av_std_parameters}. For any future work based on the results presented here, we recommend using the parameters presented in these two tables.


\section{Estimating final helium core masses}\label{sec:Hecore}

\begin{figure}
\begin{center}
\includegraphics[width=\linewidth]{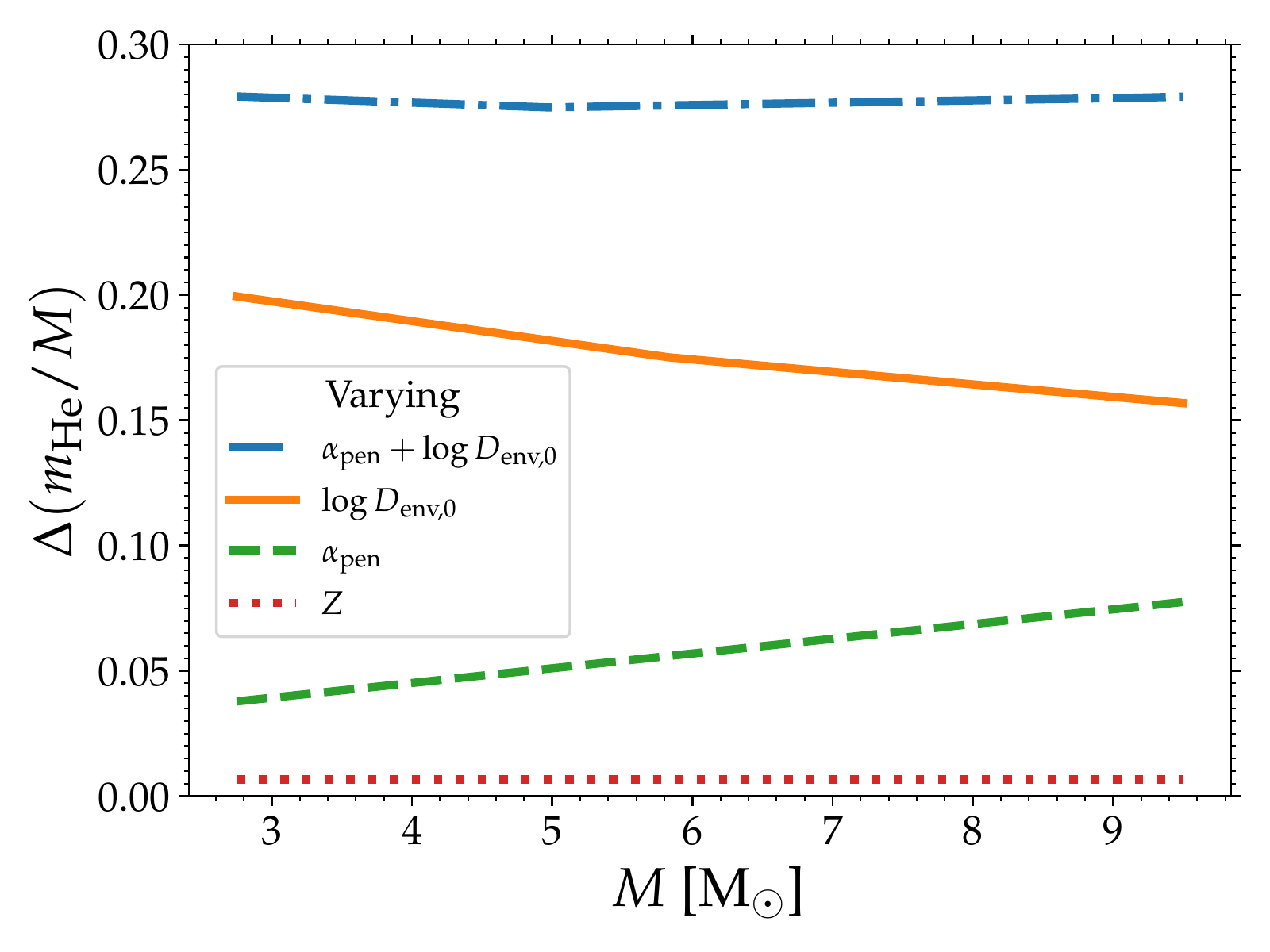}
\caption{Illustration of the effect of varying the CBM, envelope mixing, and metallicity on the predicted helium core masses as a function of stellar mass. }
\label{fig:HeCore_diff}
\end{center}
\end{figure}

\noindent The mixing history of the stars on the main-sequence can significantly impact the helium core masses.
Figure~\ref{fig:HeCore_diff} illustrates the differences in $m_{\rm He}$ obtained as a function of initial stellar mass between cases of minimum and maximum internal mixing and metallicity, using the $\boldsymbol{\psi}_7$ grid as an example. The dashed green line shows the effect on $m_{\rm He}$ when the CBM parameter $\alpha_{\rm pen}$ is changed from 0.1 to 0.4, and all other parameters are held constant ($\log D_{\rm env,0} = 1$), i.e. in this case we have $\Delta (m_{\rm He}/M) = [m_{\rm He}/M]_{\rm \alpha_{\rm pen} = 0.4} - [m_{\rm He}/M]_{\rm \alpha_{\rm pen} = 0.1}$. As seen in the figure, the differences in $m_{\rm He}$ increase for increasing stellar mass which is caused by the general increase in the convective core size with stellar mass.

The differences arising from changing $\log D_{\rm env,0} = 1$ to $6$ while fixing $\alpha_{\rm pen} = 0.1$ are shown by the full orange curve. In this case, we see that the overall effect of the envelope mixing on increasing the $m_{\rm He}$ is higher than the effect of increasing the size of the CBM region, and that the differences in $m_{\rm He}$ decrease for increasing stellar mass. This is because for the higher mass stars, less time is spent on the main-sequence and higher diffusive mixing coefficients would be needed to obtain the same change in $m_{\rm He}$. The combined effect of increasing both $\alpha_{\rm pen}$ and $\log D_{\rm env,0}$ from 0.1 and 1 to 0.4 and 6, respectively, are shown by the dot-dashed blue curve. We see that the differences in $m_{\rm He}$ between these two scenarios of minimum and maximum internal mixing are nearly constant as a function of stellar mass, and the highest differences in $m_{\rm He}$ are obtained. In contrast, varying the metal mass fraction $Z$ from 0.001 to 0.04 causes only a minor change in the $m_{\rm He}$ values obtained at the end of the main-sequence evolution, cf. dotted pink curve in Fig.~\ref{fig:HeCore_diff}.

In the addition to the quantities derived by statistical models by \cite{Pedersen2021}, we here calculate a statistical model to predict what the helium core mass $m_{\rm He}$ will be at the end of the main-sequence evolution for the 26 SPB stars based on their estimated $\boldsymbol{\theta}$ parameters. In this case we write the statistical model as

\begin{equation}
m_{{\rm He}, j} = \boldsymbol{x}_{j}^\top \boldsymbol{\beta},
	\label{Eq:stat_mod_mHe}
\end{equation}

\noindent where $m_{{\rm He}, j}$ is the helium core mass at the $j$th grid point, $\boldsymbol{x}_{j}$ is a list of fractional polynomials built from the components of $(M,Z, f_{\rm ov}\ {\rm or}\ \alpha_{\rm pen}, \log D_{\rm env,0})$, and $\boldsymbol{\beta}$ are once again the regression coefficients. The parameter $f_{\rm rot}$ has no impact on the estimated $m_{\rm He}$ as $f_{\rm rot}$ is not included in the \texttt{MESA} computations, while $X_{\rm c}/X_{\rm ini}$ is held fixed at 0.01. A comparison between original $m_{\rm He}$ values from the grids and those predicted from the statistical model is provided in Appendix~\ref{Sec:Grid_vs_stat}, while the definition of $m_{\rm He}$ in \texttt{MESA} is provided in Appendix~\ref{Sec:HeCoreDef}. The derived $m_{\rm He}$ values and their errors are provided in Table~\ref{Tab:theta_parameters_extra_averages} in Appendix~\ref{Sec:theta_estimates}.

\begin{figure}
\begin{center}
\includegraphics[width=1\linewidth]{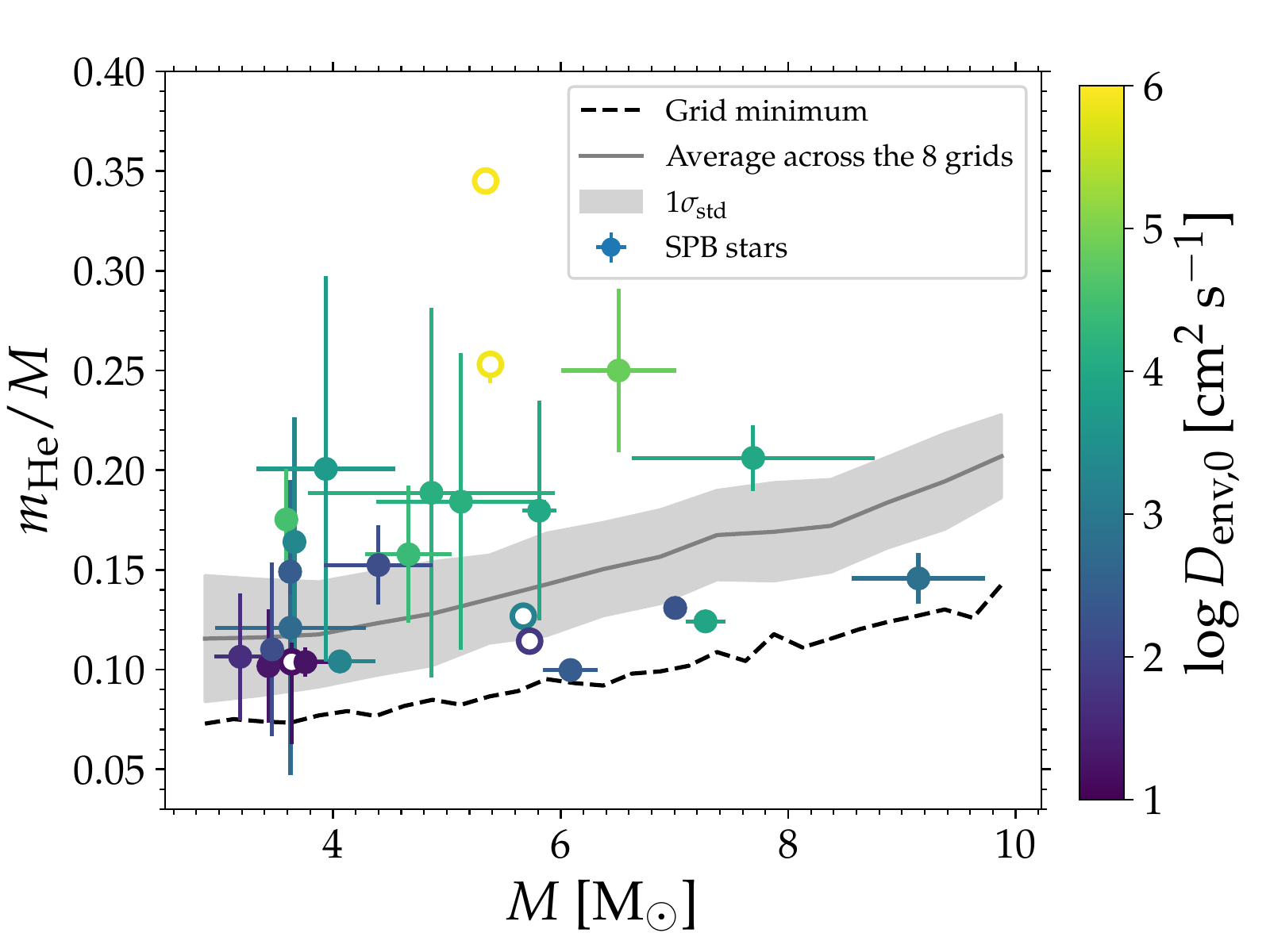}
\caption{Fractional helium core mass obtained at the end of the main-sequence evolution shown as a function of stellar mass for the 26~SPB stars modeled by \cite{Pedersen2021}. Brighter colored data points corresponds to a higher value of estimated envelope mixing. The black dashed line indicates the minimum helium core masses calculate for the $\boldsymbol{\psi}_1, \dots, \boldsymbol{\psi}_8$ grids of stellar models. The full grey line is the average helium core mass obtained across the eight grids, and the grey shaded region indicates its standard deviation.}
\label{fig:HeCore}
\end{center}
\end{figure}

Figure~\ref{fig:HeCore} shows the estimated helium core masses expected at the end of the main-sequence evolution for the 26~SPB stars as a function of their initial stellar mass. Full symbols mark the stars for which at least one other grid had an AICc distribution within $1\sigma_{\rm std}$ of the 86th percentile of the best matching model grid (values in Table~\ref{Tab:theta_parameters_extra_averages}). For the other five stars indicated by the open circles, the values from Table~\ref{Tab:theta_parameters_extra} were used. The black dashed line shows the minimum values of $m_{\rm He}$ obtained across the eight grids of stellar models, while the full grey line and the grey shaded region corresponds to the average and its $1\sigma_{\rm std}$, respectively. The color of the data points indicate the amount of envelope mixing estimated for the stars. The internal mixing of the stars causes the helium core masses to be higher than the minimum values predicted from the eight grids corresponding to the smallest amount of mixing included in the models. This is especially clear for the stars with the highest amount of envelope mixing, where the helium core masses have at least doubled in value compared to the minimum values. Consequently, not including internal mixing when calculating stellar structure and evolution models leads to underestimated helium core masses for these stars.

\cite{Kaiser2020} recently performed a detailed study of the effects of CBM on the stellar structure and evolution of $15-25$\,M$_\odot$ stars. They found relative differences of up to 70\% on the predicted core masses for stars in their considered mass range from the inclusion of extensive CBM in their 1D stellar models. In line with these results, \cite{Johnston2021} recently compiled a list of known main-sequence convective core masses for 110 stars in the mass range $1.14-24$\,M$_\odot$ derived from asteroseismology or the modeling of eclipsing binaries, showing that the standard stellar evolution models without chemical mixing frequently underestimate the core masses of the stars. The results presented here are in full agreement with \cite{Johnston2021}, who considered a much broader range in mass.

\section{Conclusions}\label{sec:Conclusions}

\noindent In this work we revisited and improved the asteroseismic modeling results and recently derived internal mixing profiles of 26~SPB stars \citep{Pedersen2021}. We find that the capability of an observed period spacing pattern to differentiate between different mixing profiles is very star dependent. For five of the stars, the mixing profiles from \cite{Pedersen2021} were determined unambiguously, whereas for the remaining stars at least one other profile is able to match the observations equally well. Therefore, we expect that these stars would benefit the most from an asteroseismic modeling approach that computes dense grids of stellar models and their theoretical oscillation properties, instead of relying on grid refinement from statistical models. Accounting for the number of mixing profiles which match the observations equally well within $1\sigma_{\rm std}^*$, we find that $\approx$55\% of the stars prefer CBM from convective penetration and $\approx$45\% from exponential core overshoot. For the envelope mixing profile, the corresponding numbers are in order of highest to lowest preference: $\approx$39\% preference for envelope mixing due to vertical shear instabilities, followed by $\approx$28\% for constant mixing, $\approx$22\% for IGW mixing, and $\approx$12\% for mixing due to meridional circulation combined with shear instabilities. Updated parameter estimates and their errors accounting for the overlap in AICc distributions are provided in Table~\ref{Tab:theta_parameters_averages} and \ref{Tab:theta_parameters_extra_averages} of Appendix~\ref{Sec:theta_estimates}.

Finally, we computed new statistical models to predict the mass of the helium core obtained at the end of main-sequence evolution. Using the best model estimates and their updated errors, the helium core masses are found to increase with initial stellar mass as expected, and are heavily influenced by the amount of mixing present in the envelopes of the stars.


\section*{acknowledgments}
\begin{acknowledgments}
\small
The author is thankful to Conny Aerts and Lars Bildsten for providing interesting discussions and useful comments at different stages of this work. The author is likewise thankful to the anonymous referee for their comments, which improved the manuscript. This research was supported in part by the National Science Foundation under Grant No. NSF PHY-1748958. It was performed in part at Aspen Center for Physics, which is supported by National Science Foundation grant PHY-1607611, and was partially supported by a grant from the Simons Foundation. 
\end{acknowledgments}

%

\vspace{15mm}




\bibliography{references.bib}{}
\bibliographystyle{aasjournal}




\appendix


\section{Example period spacing pattern and AICc distributions}\label{Sec:KIC4930889}

\begin{figure}
\begin{center}
\includegraphics[width=0.5\linewidth]{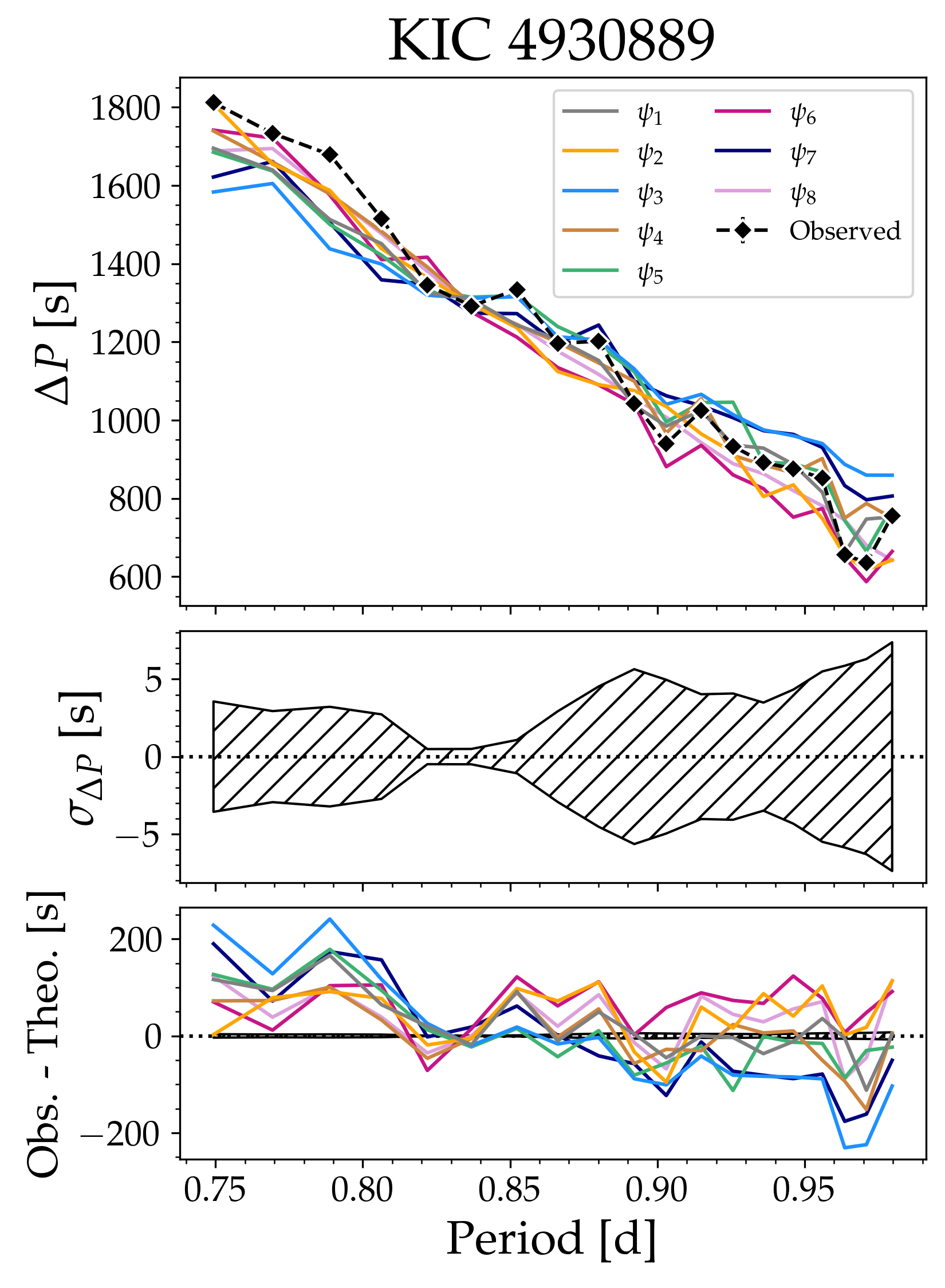}
\caption{Example observed period spacing pattern for the SPB star KIC\,4930889 compared to the best model estimates from the eight $\boldsymbol{\psi}_1, \dots, \boldsymbol{\psi}_1$ grids (top panel). The error bars are smaller than the symbol sizes. The middle panel shows the error ranges on the observed period spacing values, while the bottom panel shows the differences between the best matching theoretical pattern and the observed one for the eight grids, as indicated by the colors.}\label{fig:pattern}
\end{center}
\end{figure}

\noindent Here the SPB star KIC~4930889 is used to demonstrate what an observed period spacing pattern looks like, as well as the separations of the AICc distributions arising from the perturbations of Eq.~(\ref{Eq:stat_mod}). The observed period spacing pattern is shown in black in the top panel of Fig.~\ref{fig:pattern}, with the corresponding best matching theoretical period spacing patterns for each of the eight considered $D_{\rm mix} (r)$ are indicated by the different colors. The middle panel indicates the errors on the observed period spacing values, whereas the bottom panel gives the differences between the observed and theoretical period spacing values. As seen in the figure, these differences are larger than the observed errors.

\begin{figure*}
\begin{center}
\includegraphics[width=0.9\linewidth]{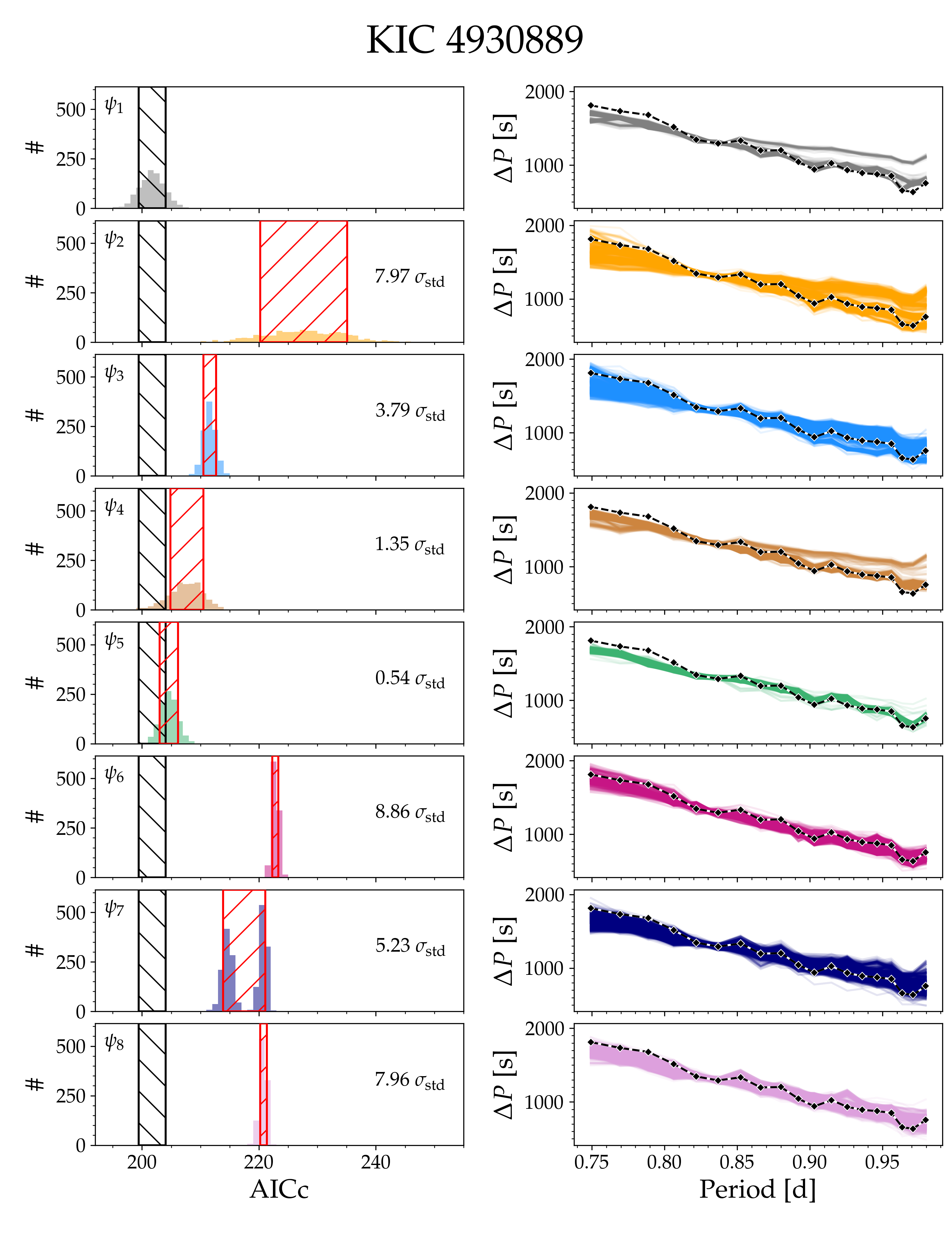}
\caption{AICc distributions of the best model estimates resulting from the 1000 perturbations of Eq.~(\ref{Eq:stat_mod}) for each of the eight $\boldsymbol{\psi}_1, \dots \boldsymbol{\psi}_8$ grids summarized in Table~\ref{Tab:Dmix} (left), and the corresponding 1000 best fitting period spacing patterns plotted against the observed period spacing pattern in black (right) for the SPB star KIC\,4930889. The black hatched region shows the 86th percentile of the distribution found to best match the observations (top panel), while the red hatched region shows the corresponding 86th percentile region for the other seven grids. The separation from the median value of the best AICc distribution and the 86th percentile regions of the other seven grids as calculated using Eq.~(\ref{Eq:sigma_std}) are indicated in each subplot.}\label{fig:example_dist}
\end{center}
\end{figure*}

Fig.~\ref{fig:example_dist} shows an example of what the AICc distributions resulting from the 1000 iterations look like for KIC\,4930889, while the corresponding 1000 best fitting period spacing patterns are shown on the right hand side for each of the eight grids and compared to the observed period spacing pattern in black. The AICc distribution for the $D_{\rm mix} (r)$ which gives the overall best match to the observation is shown in the top panel, with the 86th percentile region indicated by the black hatched area. The AICc distributions of the remaining seven grids are shown in the remaining panels, with the grid indicated in the upper left corner according to the notation in Table~\ref{Tab:Dmix}. Their 86th percentiles are indicated by their hatched red regions and their separation from the AICc distribution of the best matching model grid are indicated in each subpanel and calculated as in Eq.~(\ref{Eq:sigma_std}).

\clearpage
\section{Parameter estimates and error determination}\label{Sec:theta_estimates}

\subsection{Original error estimates from 1000 perturbations of the period spacing values}\label{Sec:error_pert}

\noindent The original reason for perturbing the $\boldsymbol{\beta}_i$ coefficients in Eq.~(\ref{Eq:stat_mod}) according to their errors and redetermining the best $\boldsymbol{\theta}$ estimates for the 26 SPB stars was to derive errors on the varied stellar parameters. As we have now increased the number of iterations from 100 to 1000, this likewise calls for a reevaluation of the errors. 

The error $\varepsilon^{\rm pert}$ arising from the 1000 perturbations is obtained by taking the full-width-at-half-maximim (FWHM) of the resulting histogram of the 1000 estimates of a given parameter (e.g. $M$ or $m_{\rm He}$), and using the resulting FWHM range in the given parameter values as the error range. All errors on the six varied stellar parameters $\left(M, Z, X_{\rm c}/X_{\rm ini}, f_{\rm ov}\ {\rm or}\ \alpha_{\rm pen}, D_{\rm env, 0}, \Omega_{\rm rot}/\Omega_{\rm crit} \right)$ in the eight model grids were computed in this way, see also \cite{Pedersen2021}. The original estimated parameters and their updated errors are provided in Table~\ref{Tab:theta_parameters}. For the additional estimated parameters ($T_{\rm eff}, \log g, \log L,  R_\star, r_{\rm cc}, m_{\rm cc}, m_{\rm He}, \log \rm{Age}$) that are not contained in $\boldsymbol{\theta}$, we included an additional error term $\varepsilon^{\rm stat}$ as discussed below. These parameters and their errors are provided in Table~\ref{Tab:theta_parameters_extra}. Here $T_{\rm eff}$, $\log g$, and $\log L$ are the asteroseismic values, $R$ is the stellar radius derived using $L = 4\pi\sigma R^2 T_{\rm eff}^4$, $m_{\rm cc}$ is the convective core mass, and $\log {\rm Age}$ is the current age of the star in Myr.

As an example, we derive the total error $\varepsilon_{m_{\rm He}}^{\rm tot}$ on the helium core masses  as a sum of the error $\varepsilon_{m_{\rm He}}^{\rm pert}$ from the 1000 perturbations of the regression coefficients in Eq.~(\ref{Eq:stat_mod}) and the error $\varepsilon_{m_{\rm He}}^{\rm stat}$ arising from the statistical model in Eq.~(\ref{Eq:stat_mod_mHe}) used to derive $m_{\rm He}$, i.e.

\begin{equation}
	\varepsilon_{m_{\rm He}}^{\rm tot} = \varepsilon_{m_{\rm He}}^{\rm pert} + \varepsilon_{m_{\rm He}}^{\rm stat}.
\end{equation}

\noindent This summation of errors is done separately for the upper and lower errors.

The error term from the statistical model $\varepsilon_{m_{\rm He}}^{\rm stat}$ is obtained by considering the deviations between the $m_{\rm He}$ values from the original eight coarser grids of \texttt{MESA} models and those derived from the statistical model as illustrated in Fig.~\ref{fig:Grid_vs_stat} in Appendix~\ref{Sec:Grid_vs_stat}. The black dashed line indicates where the $m_{\rm He}$ values from the statistical models are exactly the same as those coming directly from the \texttt{MESA} models. The spread of the data points around this line indicates the error on $m_{\rm He}$ from the statistical models. As seen in Fig.~\ref{fig:Grid_vs_stat}, this spread is not uniform and thereby $m_{\rm He}$ dependent. To take this into account, we calculate the standard deviation of $\left| m_{\rm He}^{\rm Stat} - m_{\rm He}^\texttt{MESA} \right|$ separately for all the data points above and below the black dashed lines in Fig.~\ref{fig:Grid_vs_stat}, and use these as an estimate of the upper and lower values for the statistical error term $\varepsilon_{m_{\rm He}}^{\rm stat}$. These standard deviations are plotted as a function of $m_{\rm He}^{\rm Stat}$ in Fig.~\ref{fig:stat_extra_error}, where the blue and red curves correspond to the standard deviations for the data points above ($m_{\rm He}^{\rm Stat} \geq m_{\rm He}^\texttt{MESA}$) and below ($m_{\rm He}^{\rm Stat} < m_{\rm He}^\texttt{MESA}$) the black dashed line in Fig.~\ref{fig:Grid_vs_stat}, respectively. The adopted values of $\varepsilon_{m_{\rm He}}^{\rm stat}$ are then obtained by carrying out an interpolation onto these curves for the estimated model $m_{\rm He}$ values. A similar approach was taken for the other seven additional estimated parameters ($T_{\rm eff}, \log g, \log L,  R, r_{\rm cc}, m_{\rm cc}, \log \rm{Age}$) to include the errors from the statistical models.

In Table~\ref{Tab:theta_parameters} the $\boldsymbol{\theta} = (M, Z, f_{\rm ov}\ \rm{or}\ \alpha_{\rm pen}, \log D_{\rm env,0}, X_{\rm c}/X_{\rm ini}, \Omega_{\rm rot}/\Omega_{\rm crit})$ parameters from the best model estimates by \cite{Pedersen2021} are provided, including updated errors from this work resulting from the 1000 perturbations of Eq.~(\ref{Eq:stat_mod}). The additional estimated asteroseismic parameters $(\log T_{\rm eff}, \log g, \log L, R, r_{\rm cc}/R, m_{\rm cc}/M, m_{\rm He}/M, \log {\rm Age})$ inferred from the estimated $\boldsymbol{\theta}$ parameters are provided in Table~\ref{Tab:theta_parameters_extra}. 

The error term $\varepsilon^{\rm pert}$ is essentially a measure of the robustness of the statistical model in Eq.~(\ref{Eq:stat_mod}) to predict the same theoretical period spacing values and estimate the varied parameters. As seen in Tables~\ref{Tab:theta_parameters} and \ref{Tab:theta_parameters_extra}, some of the errors on the estimated parameters turn out to be very small. This is a result of in some of the cases, the 1000 perturbations of Eq.~(\ref{Eq:stat_mod}) consistently returns (nearly) the same best model estimate for a given star. These error estimates should therefore be used with caution and instead we recommend using the parameter estimates and their errors derived and discussed in Appendix~\ref{Sec:av_std_parameters} instead.

{\movetabledown=2.05in\tabcolsep=5pt
\begin{deluxetable*}{ccccccc}
\tablecaption{Estimated $\boldsymbol{\theta}$ parameters from \cite{Pedersen2021}, including updated errors from this work. Numbers in italics indicate $f_{\rm ov}$ values. The same parameters that take into account the  modeling uncertainties discussed in Appendix~\ref{Sec:av_std_parameters} are presented in Table~\ref{Tab:theta_parameters_averages}.\label{Tab:theta_parameters}}
\tablewidth{700pt}
\tabletypesize{\small}
\tablehead{
\colhead{KIC} & \colhead{$M$} & 
\colhead{$Z$} & \colhead{$f_{\rm ov}$ or $\alpha_{\rm pen}$} & 
\colhead{$\log D_{\rm env,0}$} &
 \colhead{$X_{\rm c}/ X_{\rm ini}$} &
\colhead{$\Omega_{\rm rot}/ \Omega_{\rm crit}$}  
 \\
& \colhead{[M$_\odot$]} & & \colhead{[$H_{\rm p}$]} & \colhead{[cm$^2$\,s$^{-1}$]} &
\colhead{[\%]} & \colhead{[\%]}
} 
\startdata
		 3240411 	&	 5.341$^{+0.035}_{-0.035}$ 	&	 0.00622$^{+0.00014}_{-0.00014}$ 	&	 \emph{0.0328$^{+0.0003}_{-0.0003}$} 	&	 5.94$^{+0.01}_{-0.01}$ 	&	 68.33$^{+0.23}_{-0.23}$ 	&	 64.81$^{+0.45}_{-0.45}$\\[0.5ex]
		 3459297 	&	 3.700$^{+0.017}_{-0.017}$ 	&	 0.00664$^{+0.00006}_{-0.00006}$ 	&	 0.050$^{+0.002}_{-0.002}$ 	&	 1.72$^{+0.09}_{-0.09}$ 	&	 4.75$^{+1.43}_{-0.20}$ 	&	 52.04$^{+0.31}_{-0.31}$\\[0.5ex]
		 3865742 	&	 5.662$^{+0.059}_{-0.059}$ 	&	 0.01151$^{+0.00004}_{-0.00004}$ 	&	 0.390$^{+0.001}_{-0.001}$ 	&	 5.19$^{+0.06}_{-0.06}$ 	&	 21.93$^{+0.19}_{-0.19}$ 	&	 91.70$^{+0.23}_{-0.23}$\\[0.5ex]
		 4930889 	&	 4.375$^{+0.087}_{-0.087}$ 	&	 0.00922$^{+0.00024}_{-0.00024}$ 	&	 \emph{0.0128$^{+0.0014}_{-0.0014}$} 	&	 2.74$^{+0.19}_{-0.19}$ 	&	 36.94$^{+1.60}_{-1.60}$ 	&	 54.95$^{+0.92}_{-0.92}$\\[0.5ex]
		 4936089 	&	 3.656$^{+0.039}_{-0.039}$ 	&	 0.01533$^{+0.00030}_{-0.00030}$ 	&	 0.383$^{+0.003}_{-0.003}$ 	&	 1.07$^{+0.09}_{-0.09}$ 	&	 34.87$^{+1.97}_{-1.97}$ 	&	 10.37$^{+0.66}_{-0.66}$\\[0.5ex]
		 4939281 	&	 5.383$^{+0.014}_{-0.014}$ 	&	 0.01692$^{+0.00001}_{-0.00001}$ 	&	 0.092$^{+0.002}_{-0.002}$ 	&	 5.89$^{+0.01}_{-0.01}$ 	&	 48.01$^{+0.09}_{-0.09}$ 	&	 88.43$^{+0.13}_{-0.13}$\\[0.5ex]
		 5309849 	&	 5.673$^{+0.068}_{-0.068}$ 	&	 0.00913$^{+0.00005}_{-0.00005}$ 	&	 0.212$^{+0.003}_{-0.003}$ 	&	 3.22$^{+0.01}_{-0.01}$ 	&	 15.42$^{+0.04}_{-0.04}$ 	&	 53.99$^{+0.61}_{-0.61}$\\[0.5ex]
		 5941844 	&	 3.747$^{+0.449}_{-0.071}$ 	&	 0.01444$^{+0.00562}_{-0.00089}$ 	&	 0.211$^{+0.158}_{-0.000}$ 	&	 1.69$^{+0.13}_{-0.39}$ 	&	 93.55$^{+6.29}_{-1.71}$ 	&	 38.44$^{+1.62}_{-4.85}$\\[0.5ex]
		 6352430 	&	 3.404$^{+0.108}_{-0.036}$ 	&	 0.00885$^{+0.00015}_{-0.00015}$ 	&	 0.151$^{+0.015}_{-0.015}$ 	&	 1.61$^{+0.20}_{-0.20}$ 	&	 25.35$^{+3.34}_{-3.34}$ 	&	 55.94$^{+1.71}_{-1.71}$\\[0.5ex]
		 6462033 	&	 7.446$^{+0.048}_{-0.048}$ 	&	 0.01222$^{+0.00006}_{-0.00006}$ 	&	 0.103$^{+0.008}_{-0.038}$ 	&	 4.00$^{+0.13}_{-0.66}$ 	&	 16.79$^{+0.10}_{-0.10}$ 	&	 78.93$^{+0.63}_{-3.17}$\\[0.5ex]
		 6780397 	&	 6.286$^{+0.003}_{-0.003}$ 	&	 0.03896$^{+0.00006}_{-0.00006}$ 	&	 0.074$^{+0.001}_{-0.001}$ 	&	 1.58$^{+0.01}_{-0.01}$ 	&	 4.59$^{+0.03}_{-0.03}$ 	&	 68.92$^{+0.03}_{-0.03}$\\[0.5ex]
		 7630417 	&	 6.921$^{+0.182}_{-0.182}$ 	&	 0.01227$^{+0.00021}_{-0.00021}$ 	&	 \emph{0.0113$^{+0.0015}_{-0.0015}$} 	&	 1.80$^{+0.17}_{-0.17}$ 	&	 15.08$^{+0.64}_{-0.64}$ 	&	 59.02$^{+1.16}_{-1.16}$\\[0.5ex]
		 7760680 	&	 3.369$^{+0.112}_{-0.112}$ 	&	 0.02268$^{+0.00109}_{-0.00036}$ 	&	 0.063$^{+0.047}_{-0.016}$ 	&	 1.28$^{+0.19}_{-0.19}$ 	&	 52.68$^{+7.09}_{-2.36}$ 	&	 27.75$^{+0.84}_{-0.84}$\\[0.5ex]
		 8057661 	&	 9.520$^{+0.085}_{-0.085}$ 	&	 0.01099$^{+0.00006}_{-0.00006}$ 	&	 0.144$^{+0.004}_{-0.004}$ 	&	 3.52$^{+0.12}_{-0.12}$ 	&	 18.10$^{+0.12}_{-0.12}$ 	&	 66.88$^{+1.54}_{-1.54}$\\[0.5ex]
		 8255796 	&	 5.729$^{+0.006}_{-0.006}$ 	&	 0.01402$^{+0.00003}_{-0.00003}$ 	&	 0.158$^{+0.004}_{-0.004}$ 	&	 1.82$^{+0.01}_{-0.01}$ 	&	 2.01$^{+0.04}_{-0.04}$ 	&	 18.72$^{+0.77}_{-0.77}$\\[0.5ex]
		 8381949 	&	 6.272$^{+0.830}_{-0.166}$ 	&	 0.01228$^{+0.00432}_{-0.00029}$ 	&	 0.138$^{+0.065}_{-0.028}$ 	&	 5.76$^{+0.22}_{-0.22}$ 	&	 56.55$^{+1.94}_{-1.94}$ 	&	 81.30$^{+0.98}_{-0.98}$\\[0.5ex]
		 8459899 	&	 3.357$^{+0.137}_{-0.000}$ 	&	 0.00448$^{+0.00011}_{-0.00056}$ 	&	 0.366$^{+0.016}_{-0.078}$ 	&	 5.79$^{+0.14}_{-0.02}$ 	&	 34.25$^{+0.96}_{-6.71}$ 	&	 9.45$^{+4.29}_{-0.86}$\\[0.5ex]
		 8714886 	&	 5.847$^{+0.792}_{-0.042}$ 	&	 0.01071$^{+0.00069}_{-0.00014}$ 	&	 0.089$^{+0.013}_{-0.022}$ 	&	 3.68$^{+0.15}_{-2.31}$ 	&	 29.93$^{+0.46}_{-0.46}$ 	&	 20.15$^{+1.02}_{-1.02}$\\[0.5ex]
		 8766405 	&	 3.494$^{+0.162}_{-0.054}$ 	&	 0.00649$^{+0.00040}_{-0.00256}$ 	&	 0.331$^{+0.075}_{-0.000}$ 	&	 4.37$^{+0.47}_{-0.28}$ 	&	 21.59$^{+0.00}_{-3.24}$ 	&	 97.44$^{+1.36}_{-23.06}$\\[0.5ex]
		 9020774 	&	 3.407$^{+0.073}_{-0.219}$ 	&	 0.01045$^{+0.00107}_{-0.00537}$ 	&	 \emph{0.0120$^{+0.0047}_{-0.0079}$} 	&	 2.42$^{+3.11}_{-0.62}$ 	&	 78.04$^{+12.03}_{-26.46}$ 	&	 40.07$^{+8.51}_{-8.51}$\\[0.5ex]
		 9715425 	&	 4.715$^{+0.321}_{-0.321}$ 	&	 0.01312$^{+0.00261}_{-0.00203}$ 	&	 \emph{0.0387$^{+0.0016}_{-0.0047}$} 	&	 4.52$^{+0.48}_{-0.80}$ 	&	 48.10$^{+4.24}_{-4.24}$ 	&	 94.37$^{+0.71}_{-4.97}$\\[0.5ex]
		 10526294 	&	 3.637$^{+0.051}_{-0.153}$ 	&	 0.01156$^{+0.00371}_{-0.00074}$ 	&	 \emph{0.0228$^{+0.0045}_{-0.0197}$} 	&	 1.18$^{+0.14}_{-0.14}$ 	&	 23.29$^{+3.95}_{-3.95}$ 	&	 10.50$^{+0.53}_{-5.79}$\\[0.5ex]
		 10536147 	&	 7.503$^{+0.081}_{-0.081}$ 	&	 0.01366$^{+0.00005}_{-0.00092}$ 	&	 \emph{0.0243$^{+0.0136}_{-0.0015}$} 	&	 1.36$^{+0.23}_{-0.23}$ 	&	 97.55$^{+1.71}_{-1.71}$ 	&	 56.98$^{+18.17}_{-1.07}$\\[0.5ex]
		 11360704 	&	 4.471$^{+0.089}_{-0.089}$ 	&	 0.01461$^{+0.00030}_{-0.00030}$ 	&	 \emph{0.0394$^{+0.0015}_{-0.0015}$} 	&	 5.89$^{+0.22}_{-0.22}$ 	&	 47.17$^{+0.63}_{-0.63}$ 	&	 98.63$^{+0.30}_{-0.30}$\\[0.5ex]
		 11971405 	&	 3.673$^{+0.249}_{-0.150}$ 	&	 0.00686$^{+0.00085}_{-0.00028}$ 	&	 0.230$^{+0.043}_{-0.071}$ 	&	 4.31$^{+1.17}_{-0.91}$ 	&	 46.44$^{+4.37}_{-4.37}$ 	&	 94.91$^{+2.99}_{-0.00}$\\[0.5ex]
		 12258330 	&	 3.534$^{+0.102}_{-0.015}$ 	&	 0.00450$^{+0.00023}_{-0.00023}$ 	&	 \emph{0.0205$^{+0.0011}_{-0.0011}$} 	&	 3.50$^{+0.12}_{-0.12}$ 	&	 67.66$^{+0.96}_{-0.96}$ 	&	 81.78$^{+0.62}_{-0.62}$\\[0.5ex]
 \enddata
\end{deluxetable*}
}

{\movetabledown=2.05in\tabcolsep=5pt
\begin{deluxetable*}{ccccccccccc}
\tablecaption{Additional estimated parameters. The listed $T_{\rm eff}$, $\log g$, and $\log L$ are the asteroseismic values estimated from the asteroseismic modeling. The same parameters that take into account the modeling uncertainties discussed in Appendix~\ref{Sec:av_std_parameters} are presented in Table~\ref{Tab:theta_parameters_extra_averages}.\label{Tab:theta_parameters_extra}}
\tablewidth{700pt}
\tabletypesize{\small}
\tablehead{
\colhead{KIC} & \colhead{$\log T_{\rm eff}$} & \colhead{$\log g$}  & \colhead{$\log L$}  & \colhead{$R$}  & \colhead{$r _{\rm cc}/R$} & \colhead{$m _{\rm cc}/M$}  & \colhead{$m _{\rm He}/M$}  & \colhead{$\log$ Age}\\
&  \colhead{[K]}  & \colhead{[cm s$^{-2}$]}  & \colhead{[L$_\odot$]} & \colhead{[R$_\odot$]}
 & &&&\colhead{[Myr]}
} 
\startdata
		 3240411 	&	 4.31$^{+0.01}_{-0.01}$ 	&	 4.09$^{+0.05}_{-0.02}$ 	&	 3.25$^{+0.03}_{-0.03}$ 	&	 3.31$^{+0.18}_{-0.18}$ 	&	 0.175$^{+0.003}_{-0.005}$ 	&	 0.292$^{+0.008}_{-0.011}$ 	&	 0.345$^{+0.006}_{-0.006}$ 	&	 1.871$^{+0.026}_{-0.024}$\\[0.5ex]
		 3459297 	&	 4.11$^{+0.04}_{-0.01}$ 	&	 3.83$^{+0.06}_{-0.03}$ 	&	 2.58$^{+0.06}_{-0.06}$ 	&	 3.90$^{+0.57}_{-0.57}$ 	&	 0.050$^{+0.003}_{-0.001}$ 	&	 0.065$^{+0.001}_{-0.001}$ 	&	 0.069$^{+0.001}_{-0.001}$ 	&	 2.166$^{+0.052}_{-0.026}$\\[0.5ex]
		 3865742 	&	 4.24$^{+0.02}_{-0.02}$ 	&	 3.56$^{+0.06}_{-0.06}$ 	&	 3.56$^{+0.04}_{-0.04}$ 	&	 6.61$^{+0.63}_{-0.63}$ 	&	 0.077$^{+0.002}_{-0.002}$ 	&	 0.186$^{+0.008}_{-0.011}$ 	&	 0.235$^{+0.015}_{-0.013}$ 	&	 1.998$^{+0.042}_{-0.035}$\\[0.5ex]
		 4930889 	&	 4.18$^{+0.03}_{-0.02}$ 	&	 3.92$^{+0.05}_{-0.04}$ 	&	 2.84$^{+0.08}_{-0.05}$ 	&	 3.79$^{+0.51}_{-0.51}$ 	&	 0.097$^{+0.004}_{-0.004}$ 	&	 0.149$^{+0.008}_{-0.008}$ 	&	 0.107$^{+0.011}_{-0.010}$ 	&	 2.011$^{+0.062}_{-0.054}$\\[0.5ex]
		 4936089 	&	 4.07$^{+0.02}_{-0.02}$ 	&	 3.79$^{+0.08}_{-0.06}$ 	&	 2.46$^{+0.06}_{-0.09}$ 	&	 4.02$^{+0.60}_{-0.60}$ 	&	 0.075$^{+0.005}_{-0.004}$ 	&	 0.118$^{+0.010}_{-0.008}$ 	&	 0.102$^{+0.008}_{-0.006}$ 	&	 2.287$^{+0.050}_{-0.033}$\\[0.5ex]
		 4939281 	&	 4.22$^{+0.01}_{-0.02}$ 	&	 3.70$^{+0.05}_{-0.03}$ 	&	 3.31$^{+0.03}_{-0.04}$ 	&	 5.38$^{+0.38}_{-0.38}$ 	&	 0.105$^{+0.002}_{-0.003}$ 	&	 0.243$^{+0.008}_{-0.013}$ 	&	 0.253$^{+0.007}_{-0.009}$ 	&	 2.123$^{+0.027}_{-0.025}$\\[0.5ex]
		 5309849 	&	 4.21$^{+0.02}_{-0.02}$ 	&	 3.70$^{+0.05}_{-0.05}$ 	&	 3.28$^{+0.07}_{-0.05}$ 	&	 5.55$^{+0.75}_{-0.75}$ 	&	 0.073$^{+0.002}_{-0.003}$ 	&	 0.131$^{+0.005}_{-0.007}$ 	&	 0.127$^{+0.006}_{-0.007}$ 	&	 1.811$^{+0.038}_{-0.029}$\\[0.5ex]
		 5941844 	&	 4.15$^{+0.03}_{-0.02}$ 	&	 4.29$^{+0.11}_{-0.05}$ 	&	 2.28$^{+0.03}_{-0.09}$ 	&	 2.32$^{+0.49}_{-0.49}$ 	&	 0.163$^{+0.012}_{-0.018}$ 	&	 0.240$^{+0.027}_{-0.095}$ 	&	 0.102$^{+0.023}_{-0.006}$ 	&	 1.354$^{+0.021}_{-0.544}$\\[0.5ex]
		 6352430 	&	 4.07$^{+0.03}_{-0.03}$ 	&	 3.85$^{+0.07}_{-0.05}$ 	&	 2.35$^{+0.04}_{-0.08}$ 	&	 3.62$^{+0.60}_{-0.60}$ 	&	 0.074$^{+0.007}_{-0.007}$ 	&	 0.120$^{+0.019}_{-0.014}$ 	&	 0.096$^{+0.013}_{-0.011}$ 	&	 2.267$^{+0.113}_{-0.110}$\\[0.5ex]
		 6462033 	&	 4.26$^{+0.01}_{-0.02}$ 	&	 3.70$^{+0.09}_{-0.04}$ 	&	 3.60$^{+0.04}_{-0.04}$ 	&	 6.42$^{+0.56}_{-0.56}$ 	&	 0.080$^{+0.002}_{-0.002}$ 	&	 0.145$^{+0.004}_{-0.010}$ 	&	 0.121$^{+0.005}_{-0.007}$ 	&	 1.564$^{+0.022}_{-0.022}$\\[0.5ex]
		 6780397 	&	 4.13$^{+0.02}_{-0.02}$ 	&	 3.54$^{+0.04}_{-0.05}$ 	&	 3.17$^{+0.06}_{-0.04}$ 	&	 6.97$^{+0.75}_{-0.75}$ 	&	 0.053$^{+0.002}_{-0.002}$ 	&	 0.088$^{+0.005}_{-0.004}$ 	&	 0.083$^{+0.006}_{-0.002}$ 	&	 1.838$^{+0.026}_{-0.017}$\\[0.5ex]
		 7630417 	&	 4.24$^{+0.02}_{-0.02}$ 	&	 3.67$^{+0.06}_{-0.05}$ 	&	 3.53$^{+0.05}_{-0.04}$ 	&	 6.39$^{+0.82}_{-0.82}$ 	&	 0.073$^{+0.002}_{-0.002}$ 	&	 0.131$^{+0.008}_{-0.011}$ 	&	 0.135$^{+0.012}_{-0.014}$ 	&	 1.632$^{+0.081}_{-0.076}$\\[0.5ex]
		 7760680 	&	 4.06$^{+0.04}_{-0.03}$ 	&	 3.99$^{+0.11}_{-0.02}$ 	&	 2.12$^{+0.21}_{-0.13}$ 	&	 2.94$^{+1.09}_{-0.89}$ 	&	 0.113$^{+0.015}_{-0.025}$ 	&	 0.184$^{+0.033}_{-0.066}$ 	&	 0.074$^{+0.009}_{-0.009}$ 	&	 2.291$^{+0.120}_{-0.117}$\\[0.5ex]
		 8057661 	&	 4.33$^{+0.01}_{-0.01}$ 	&	 3.77$^{+0.05}_{-0.03}$ 	&	 3.96$^{+0.04}_{-0.06}$ 	&	 6.81$^{+0.50}_{-0.50}$ 	&	 0.099$^{+0.002}_{-0.003}$ 	&	 0.185$^{+0.009}_{-0.012}$ 	&	 0.156$^{+0.006}_{-0.006}$ 	&	 1.356$^{+0.027}_{-0.024}$\\[0.5ex]
		 8255796 	&	 4.19$^{+0.01}_{-0.02}$ 	&	 3.64$^{+0.05}_{-0.03}$ 	&	 3.25$^{+0.03}_{-0.04}$ 	&	 5.93$^{+0.47}_{-0.47}$ 	&	 0.051$^{+0.002}_{-0.002}$ 	&	 0.085$^{+0.008}_{-0.006}$ 	&	 0.114$^{+0.005}_{-0.003}$ 	&	 1.813$^{+0.024}_{-0.021}$\\[0.5ex]
		 8381949 	&	 4.31$^{+0.04}_{-0.02}$ 	&	 3.96$^{+0.06}_{-0.10}$ 	&	 3.47$^{+0.23}_{-0.05}$ 	&	 4.29$^{+1.42}_{-0.98}$ 	&	 0.143$^{+0.005}_{-0.005}$ 	&	 0.244$^{+0.011}_{-0.011}$ 	&	 0.229$^{+0.043}_{-0.028}$ 	&	 1.747$^{+0.046}_{-0.096}$\\[0.5ex]
		 8459899 	&	 4.23$^{+0.02}_{-0.02}$ 	&	 3.84$^{+0.06}_{-0.06}$ 	&	 2.98$^{+0.17}_{-0.06}$ 	&	 3.63$^{+1.00}_{-0.64}$ 	&	 0.111$^{+0.004}_{-0.005}$ 	&	 0.250$^{+0.028}_{-0.021}$ 	&	 0.332$^{+0.021}_{-0.026}$ 	&	 2.724$^{+0.046}_{-0.068}$\\[0.5ex]
		 8714886 	&	 4.25$^{+0.02}_{-0.03}$ 	&	 3.85$^{+0.06}_{-0.04}$ 	&	 3.30$^{+0.11}_{-0.04}$ 	&	 4.74$^{+1.05}_{-0.75}$ 	&	 0.095$^{+0.003}_{-0.005}$ 	&	 0.148$^{+0.005}_{-0.009}$ 	&	 0.100$^{+0.024}_{-0.005}$ 	&	 1.732$^{+0.036}_{-0.173}$\\[0.5ex]
		 8766405 	&	 4.13$^{+0.04}_{-0.02}$ 	&	 3.45$^{+0.26}_{-0.06}$ 	&	 3.00$^{+0.06}_{-0.06}$ 	&	 5.80$^{+0.95}_{-1.66}$ 	&	 0.057$^{+0.010}_{-0.002}$ 	&	 0.158$^{+0.040}_{-0.017}$ 	&	 0.189$^{+0.044}_{-0.014}$ 	&	 2.483$^{+0.117}_{-0.162}$\\[0.5ex]
		 9020774 	&	 4.14$^{+0.02}_{-0.02}$ 	&	 4.26$^{+0.14}_{-0.08}$ 	&	 2.24$^{+0.25}_{-0.15}$ 	&	 2.27$^{+0.63}_{-0.98}$ 	&	 0.145$^{+0.024}_{-0.049}$ 	&	 0.205$^{+0.057}_{-0.016}$ 	&	 0.094$^{+0.074}_{-0.012}$ 	&	 2.016$^{+0.294}_{-0.109}$\\[0.5ex]
		 9715425 	&	 4.21$^{+0.03}_{-0.02}$ 	&	 3.86$^{+0.08}_{-0.04}$ 	&	 3.04$^{+0.08}_{-0.05}$ 	&	 4.27$^{+0.64}_{-0.64}$ 	&	 0.107$^{+0.004}_{-0.004}$ 	&	 0.194$^{+0.009}_{-0.018}$ 	&	 0.205$^{+0.035}_{-0.035}$ 	&	 2.102$^{+0.069}_{-0.147}$\\[0.5ex]
		 10526294 	&	 4.08$^{+0.04}_{-0.03}$ 	&	 3.74$^{+0.13}_{-0.07}$ 	&	 2.54$^{+0.04}_{-0.16}$ 	&	 4.37$^{+1.01}_{-1.23}$ 	&	 0.065$^{+0.006}_{-0.005}$ 	&	 0.114$^{+0.012}_{-0.022}$ 	&	 0.104$^{+0.010}_{-0.041}$ 	&	 2.262$^{+0.100}_{-0.088}$\\[0.5ex]
		 10536147 	&	 4.34$^{+0.01}_{-0.01}$ 	&	 4.24$^{+0.05}_{-0.03}$ 	&	 3.41$^{+0.03}_{-0.07}$ 	&	 3.45$^{+0.75}_{-0.28}$ 	&	 0.208$^{+0.007}_{-0.008}$ 	&	 0.289$^{+0.016}_{-0.012}$ 	&	 0.173$^{+0.036}_{-0.016}$ 	&	 0.267$^{+1.289}_{-0.117}$\\[0.5ex]
		 11360704 	&	 4.24$^{+0.01}_{-0.02}$ 	&	 3.80$^{+0.05}_{-0.03}$ 	&	 3.17$^{+0.06}_{-0.04}$ 	&	 4.25$^{+0.44}_{-0.44}$ 	&	 0.125$^{+0.004}_{-0.004}$ 	&	 0.269$^{+0.012}_{-0.012}$ 	&	 0.312$^{+0.021}_{-0.012}$ 	&	 2.328$^{+0.061}_{-0.049}$\\[0.5ex]
		 11971405 	&	 4.18$^{+0.03}_{-0.02}$ 	&	 3.99$^{+0.05}_{-0.01}$ 	&	 2.69$^{+0.07}_{-0.06}$ 	&	 3.14$^{+0.53}_{-0.43}$ 	&	 0.110$^{+0.005}_{-0.007}$ 	&	 0.168$^{+0.010}_{-0.027}$ 	&	 0.150$^{+0.013}_{-0.049}$ 	&	 2.283$^{+0.183}_{-0.233}$\\[0.5ex]
		 12258330 	&	 4.20$^{+0.02}_{-0.02}$ 	&	 4.26$^{+0.03}_{-0.03}$ 	&	 2.48$^{+0.04}_{-0.07}$ 	&	 2.30$^{+0.30}_{-0.28}$ 	&	 0.144$^{+0.006}_{-0.004}$ 	&	 0.195$^{+0.014}_{-0.009}$ 	&	 0.138$^{+0.007}_{-0.009}$ 	&	 2.074$^{+0.046}_{-0.112}$\\[0.5ex]
 \enddata
\end{deluxetable*}
}

\clearpage
\subsection{Stellar parameter comparisons and updated errors}\label{Sec:av_std_parameters}

\begin{figure*}
\begin{center}
\includegraphics[width=1\linewidth]{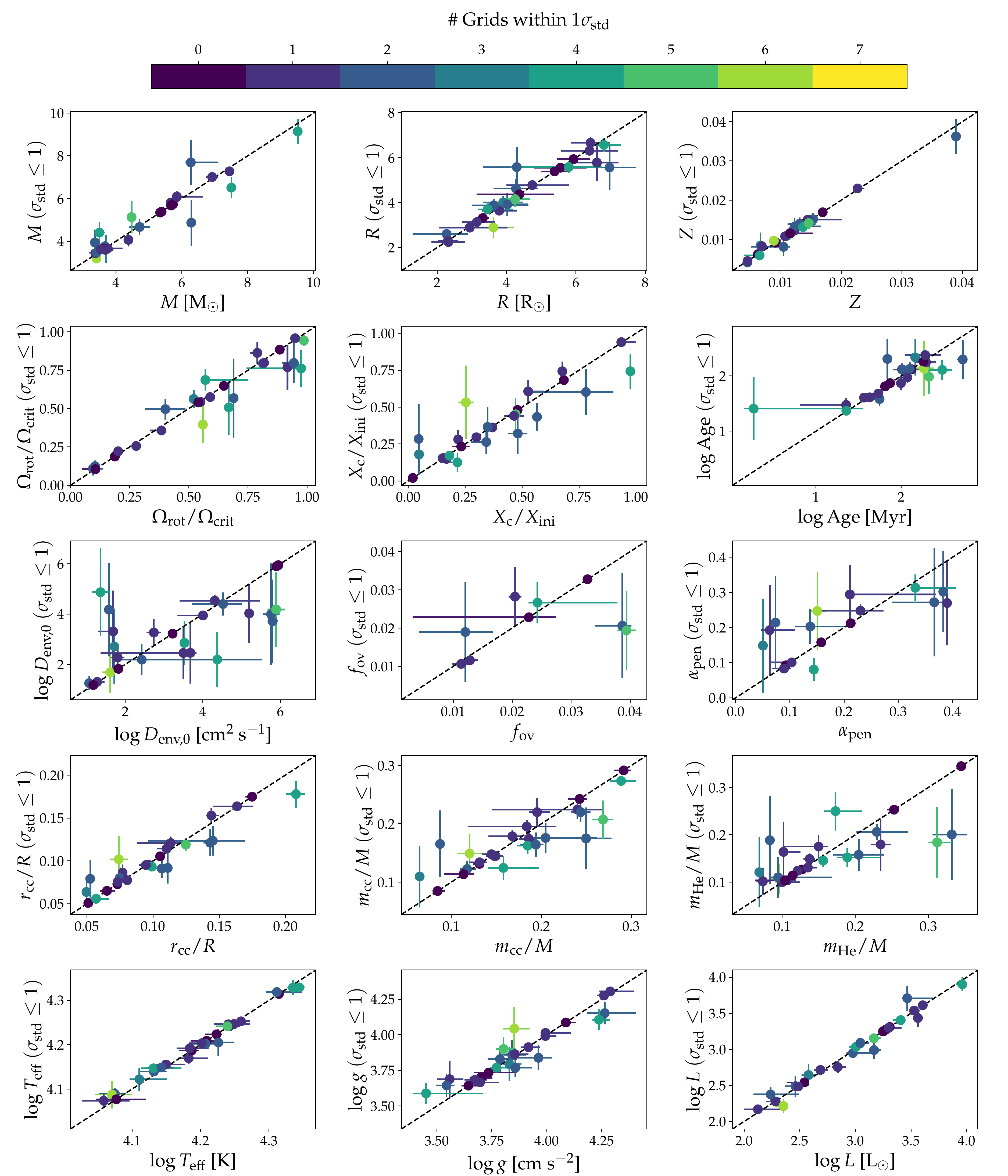}
\caption{Comparison of the average parameter estimates obtained for all grids for which their 86th percentile is separated from the median of the AICc distribution of the best matching model by less than $1\sigma_{\rm std}$ (y-axis), and the parameters from the best matching model (x-axis) for the 26~SPB stars. The colors indicate the number of grids with $\# \sigma_{\rm std} \leq 1$, i.e. the number grids in addition to the best model estimate which were included in the calculation of the average parameter value.}\label{fig:para_comp}
\end{center}
\end{figure*}

\noindent Aside from the errors on the estimated parameters resulting from the 1000 perturbations, our inability to distinguish between the best matching models and those with AICc distributions which overlap significantly with the one of our best model estimate likewise give rise to errors on the estimated parameters. For these other model grids, the estimated $\boldsymbol{\theta}$ values essentially match the observations just as well as the ones from the grid with the best model estimate. To quantify this, we calculate the average of the estimated $\boldsymbol{\theta}$ and their standard deviation for the model grids where the 86th percentile of the AICc distribution is less than $1\sigma_{\rm std}^*$ away from the median of the distribution for the best model estimate. These averages and their $1\sigma$ standard deviations are plotted against the $\boldsymbol{\theta}$ from the best model estimate in Fig.~\ref{fig:para_comp}. The data points are color coded according to the number of grids for which $\# \sigma_{\rm std} \leq 1$. For the stars where an exponentially decreasing diffusive overshoot mixing profile was preferred in the CBM region and for which an AICc distribution from a grid with convective penetration obtained a $\# \sigma_{\rm std} \leq 1$, the average $f_{\rm ov}$ is calculated by using the conversion $\alpha_{\rm pen} \approx 10 f_{\rm ov}$. The same is the case for the opposite scenario where convective penetration provides the best match to the observed period spacing pattern, but grids with exponential diffusive overshoot obtaining $\# \sigma_{\rm std} \leq 1$ are found for the same star.

We find that in general these averages are very similar if not identical to the values of the best model estimate from \cite{Pedersen2021} within their errors. The largest differences are obtained for $\log D_{\rm env,0}$ and $m_{\rm He}$, which both show the same general trend that when the best model estimate gives a lower $\log D_{\rm env,0}$ and/or $m_{\rm He}$ value then their corresponding averages are higher. Higher values of $\log D_{\rm env,0}$ and/or $m_{\rm He}$ have lower corresponding averages for the models with AICc distributions within $1\sigma_{\rm std}$. 

We provide the calculated averages and their standard deviations for the 21 SPB stars which have at least one other grid separated from the AICc distribution of the overall best matching model grid by $\leq 1\sigma_{\rm std} $ in Tables~\ref{Tab:theta_parameters_averages} and \ref{Tab:theta_parameters_extra_averages}.


{\movetabledown=2.05in\tabcolsep=5pt
\begin{deluxetable*}{ccccccc}
\tablecaption{Calculated averages of the estimated $\boldsymbol{\theta}$ parameters and their standard deviations for all grids with $\# \sigma_{\rm std} \leq 1$. The five empty entries corresponds to the five stars, where none of the other seven grids were separated from the AICc distribution of the best matching model by $\leq 1\sigma_{\rm std}$.  Numbers in italics indicate $f_{\rm ov}$ values. \label{Tab:theta_parameters_averages}}
\tablewidth{700pt}
\tabletypesize{\small}
\tablehead{
\colhead{KIC} & \colhead{$M$} & 
\colhead{$Z$} & \colhead{$f_{\rm ov}$ or $\alpha_{\rm pen}$} & 
\colhead{$\log D_{\rm env,0}$} &
 \colhead{$X_{\rm c}/ X_{\rm ini}$} &
\colhead{$\Omega_{\rm rot}/ \Omega_{\rm crit}$}  
 \\
& \colhead{[M$_\odot$]} & & \colhead{[$H_{\rm p}$]} & \colhead{[cm$^2$\,s$^{-1}$]} &
\colhead{[\%]} & \colhead{[\%]}
} 
\startdata
		 3240411 	&	 - 	&	 - 	&	 - 	&	 - 	&	 - 	&	 -\\[0.5ex]
		 3459297 	&	 3.63$\pm$0.66 	&	 0.00845$\pm$0.00332 	&	 0.15$\pm$0.13 	&	 2.7$\pm$1.5 	&	 18.0$\pm$13.9 	&	 56.3$\pm$6.0\\[0.5ex]
		 3865742 	&	 5.81$\pm$0.15 	&	 0.01159$\pm$0.00008 	&	 0.27$\pm$0.12 	&	 4.0$\pm$1.2 	&	 28.1$\pm$6.2 	&	 76.9$\pm$14.8\\[0.5ex]
		 4930889 	&	 4.06$\pm$0.31 	&	 0.00924$\pm$0.00002 	&	 \emph{0.012$\pm$0.001} 	&	 3.3$\pm$0.5 	&	 36.2$\pm$0.7 	&	 54.4$\pm$0.6\\[0.5ex]
		 4936089 	&	 3.76$\pm$0.24 	&	 0.01512$\pm$0.00169 	&	 0.30$\pm$0.11 	&	 1.3$\pm$0.3 	&	 36.6$\pm$13.4 	&	 12.5$\pm$3.8\\[0.5ex]
		 4939281 	&	 - 	&	 - 	&	 - 	&	 - 	&	 - 	&	 -\\[0.5ex]
		 5309849 	&	 - 	&	 - 	&	 - 	&	 - 	&	 - 	&	 -\\[0.5ex]
		 5941844 	&	 3.66$\pm$0.09 	&	 0.01501$\pm$0.00057 	&	 0.29$\pm$0.08 	&	 3.3$\pm$1.6 	&	 94.0$\pm$0.4 	&	 35.7$\pm$2.7\\[0.5ex]
		 6352430 	&	 3.18$\pm$0.23 	&	 0.00964$\pm$0.00093 	&	 0.25$\pm$0.11 	&	 1.7$\pm$0.8 	&	 53.2$\pm$24.9 	&	 39.6$\pm$11.8\\[0.5ex]
		 6462033 	&	 7.27$\pm$0.17 	&	 0.01287$\pm$0.00065 	&	 0.10$\pm$0.00 	&	 3.9$\pm$0.1 	&	 14.9$\pm$1.9 	&	 86.3$\pm$7.4\\[0.5ex]
		 6780397 	&	 4.87$\pm$1.08 	&	 0.03625$\pm$0.00445 	&	 0.21$\pm$0.13 	&	 4.2$\pm$1.9 	&	 28.4$\pm$23.8 	&	 56.8$\pm$25.7\\[0.5ex]
		 7630417 	&	 7.01$\pm$0.09 	&	 0.01226$\pm$0.00001 	&	 \emph{0.011$\pm$0.001} 	&	 2.3$\pm$0.5 	&	 15.3$\pm$0.2 	&	 57.5$\pm$1.5\\[0.5ex]
		 7760680 	&	 3.43$\pm$0.06 	&	 0.02299$\pm$0.00031 	&	 0.19$\pm$0.13 	&	 1.3$\pm$0.0 	&	 60.6$\pm$7.9 	&	 25.6$\pm$2.1\\[0.5ex]
		 8057661 	&	 9.14$\pm$0.59 	&	 0.01091$\pm$0.00039 	&	 0.08$\pm$0.03 	&	 2.9$\pm$0.9 	&	 17.1$\pm$1.7 	&	 50.8$\pm$17.7\\[0.5ex]
		 8255796 	&	 - 	&	 - 	&	 - 	&	 - 	&	 - 	&	 -\\[0.5ex]
		 8381949 	&	 7.69$\pm$1.06 	&	 0.01330$\pm$0.00209 	&	 0.20$\pm$0.05 	&	 4.0$\pm$2.0 	&	 43.3$\pm$9.3 	&	 79.4$\pm$2.5\\[0.5ex]
		 8459899 	&	 3.94$\pm$0.61 	&	 0.00412$\pm$0.00026 	&	 0.27$\pm$0.15 	&	 3.7$\pm$1.7 	&	 26.4$\pm$7.9 	&	 10.8$\pm$4.5\\[0.5ex]
		 8714886 	&	 6.09$\pm$0.24 	&	 0.01079$\pm$0.00008 	&	 0.08$\pm$0.01 	&	 2.5$\pm$1.2 	&	 29.6$\pm$0.4 	&	 22.2$\pm$2.0\\[0.5ex]
		 8766405 	&	 4.40$\pm$0.48 	&	 0.00591$\pm$0.00092 	&	 0.31$\pm$0.04 	&	 2.2$\pm$1.1 	&	 12.6$\pm$6.5 	&	 76.2$\pm$12.0\\[0.5ex]
		 9020774 	&	 3.47$\pm$0.04 	&	 0.00810$\pm$0.00232 	&	 \emph{0.019$\pm$0.013} 	&	 2.2$\pm$0.6 	&	 60.2$\pm$15.6 	&	 49.6$\pm$6.8\\[0.5ex]
		 9715425 	&	 4.66$\pm$0.38 	&	 0.01398$\pm$0.00158 	&	 \emph{0.021$\pm$0.014} 	&	 4.4$\pm$0.5 	&	 32.1$\pm$13.6 	&	 79.8$\pm$12.9\\[0.5ex]
		 10526294 	&	 - 	&	 - 	&	 - 	&	 - 	&	 - 	&	 -\\[0.5ex]
		 10536147 	&	 6.51$\pm$0.51 	&	 0.01314$\pm$0.00033 	&	 \emph{0.027$\pm$0.005} 	&	 4.9$\pm$1.8 	&	 74.3$\pm$11.7 	&	 68.6$\pm$6.9\\[0.5ex]
		 11360704 	&	 5.13$\pm$0.75 	&	 0.01418$\pm$0.00160 	&	 \emph{0.019$\pm$0.010} 	&	 4.2$\pm$1.5 	&	 45.1$\pm$10.8 	&	 94.3$\pm$3.8\\[0.5ex]
		 11971405 	&	 3.59$\pm$0.08 	&	 0.00826$\pm$0.00140 	&	 0.25$\pm$0.02 	&	 4.5$\pm$0.2 	&	 44.1$\pm$2.3 	&	 95.9$\pm$0.9\\[0.5ex]
		 12258330 	&	 3.62$\pm$0.09 	&	 0.00448$\pm$0.00002 	&	 \emph{0.028$\pm$0.008} 	&	 2.5$\pm$1.0 	&	 74.3$\pm$6.6 	&	 79.8$\pm$1.9\\[0.5ex]
 \enddata
\end{deluxetable*}
}

{\movetabledown=2.05in\tabcolsep=5pt
\begin{deluxetable*}{ccccccccccc}
\tablecaption{Calculated averages of the additional estimated parameters and their standard deviations for all grids with $\# \sigma_{\rm std} \leq 1$. The five empty entries corresponds to the five stars, where none of the other seven grids were separated from the AICc distribution of the best matching model by $\leq 1\sigma_{\rm std}$. The listed $T_{\rm eff}$, $\log g$, and $\log L$ are the asteroseismic values estimated from the asteroseismic modeling.\label{Tab:theta_parameters_extra_averages}}
\tablewidth{700pt}
\tabletypesize{\small}
\tablehead{
\colhead{KIC} & \colhead{$\log T_{\rm eff}$} & \colhead{$\log g$}  & \colhead{$\log L$}  & \colhead{$R$}  & \colhead{$r _{\rm cc}/R$} & \colhead{$m _{\rm cc}/M$}  & \colhead{$m _{\rm He}/M$}  & \colhead{$\log$ Age}\\
&  \colhead{[K]}  & \colhead{[cm s$^{-2}$]}  & \colhead{[L$_\odot$]} & \colhead{[R$_\odot$]}
 & &&&\colhead{[Myr]}
} 
\startdata
		 3240411 	&	 - 	&	 - 	&	 - 	&	 - 	&	 - 	&	 - 	&	 - 	&	 -\\[0.5ex]
		 3459297 	&	 4.122$\pm$0.026 	&	 3.794$\pm$0.119 	&	 2.644$\pm$0.149 	&	 4.00$\pm$0.29 	&	 0.064$\pm$0.019 	&	 0.110$\pm$0.053 	&	 0.1210$\pm$0.0741 	&	 2.34$\pm$0.33\\[0.5ex]
		 3865742 	&	 4.241$\pm$0.001 	&	 3.689$\pm$0.130 	&	 3.434$\pm$0.127 	&	 5.78$\pm$0.83 	&	 0.086$\pm$0.009 	&	 0.174$\pm$0.012 	&	 0.1797$\pm$0.0550 	&	 1.87$\pm$0.13\\[0.5ex]
		 4930889 	&	 4.170$\pm$0.013 	&	 3.913$\pm$0.004 	&	 2.754$\pm$0.088 	&	 3.64$\pm$0.15 	&	 0.095$\pm$0.002 	&	 0.145$\pm$0.004 	&	 0.1041$\pm$0.0031 	&	 2.12$\pm$0.11\\[0.5ex]
		 4936089 	&	 4.089$\pm$0.013 	&	 3.829$\pm$0.090 	&	 2.490$\pm$0.146 	&	 3.92$\pm$0.51 	&	 0.080$\pm$0.013 	&	 0.123$\pm$0.014 	&	 0.1037$\pm$0.0073 	&	 2.22$\pm$0.08\\[0.5ex]
		 4939281 	&	 - 	&	 - 	&	 - 	&	 - 	&	 - 	&	 - 	&	 - 	&	 -\\[0.5ex]
		 5309849 	&	 - 	&	 - 	&	 - 	&	 - 	&	 - 	&	 - 	&	 - 	&	 -\\[0.5ex]
		 5941844 	&	 4.155$\pm$0.006 	&	 4.304$\pm$0.016 	&	 2.277$\pm$0.006 	&	 2.25$\pm$0.07 	&	 0.164$\pm$0.000 	&	 0.224$\pm$0.016 	&	 0.1640$\pm$0.0624 	&	 1.48$\pm$0.12\\[0.5ex]
		 6352430 	&	 4.087$\pm$0.031 	&	 4.043$\pm$0.150 	&	 2.213$\pm$0.103 	&	 2.89$\pm$0.48 	&	 0.102$\pm$0.027 	&	 0.149$\pm$0.033 	&	 0.1064$\pm$0.0317 	&	 2.14$\pm$0.50\\[0.5ex]
		 6462033 	&	 4.253$\pm$0.006 	&	 3.666$\pm$0.030 	&	 3.611$\pm$0.006 	&	 6.66$\pm$0.24 	&	 0.078$\pm$0.003 	&	 0.148$\pm$0.003 	&	 0.1242$\pm$0.0030 	&	 1.61$\pm$0.04\\[0.5ex]
		 6780397 	&	 4.140$\pm$0.010 	&	 3.644$\pm$0.083 	&	 2.989$\pm$0.128 	&	 5.56$\pm$1.00 	&	 0.079$\pm$0.022 	&	 0.166$\pm$0.057 	&	 0.1887$\pm$0.0927 	&	 2.31$\pm$0.37\\[0.5ex]
		 7630417 	&	 4.246$\pm$0.005 	&	 3.684$\pm$0.011 	&	 3.537$\pm$0.010 	&	 6.31$\pm$0.08 	&	 0.076$\pm$0.002 	&	 0.134$\pm$0.003 	&	 0.1309$\pm$0.0042 	&	 1.61$\pm$0.02\\[0.5ex]
		 7760680 	&	 4.074$\pm$0.015 	&	 4.013$\pm$0.019 	&	 2.168$\pm$0.045 	&	 2.89$\pm$0.05 	&	 0.121$\pm$0.007 	&	 0.195$\pm$0.011 	&	 0.1018$\pm$0.0283 	&	 2.25$\pm$0.04\\[0.5ex]
		 8057661 	&	 4.329$\pm$0.017 	&	 3.769$\pm$0.005 	&	 3.901$\pm$0.097 	&	 6.57$\pm$0.30 	&	 0.093$\pm$0.004 	&	 0.163$\pm$0.015 	&	 0.1457$\pm$0.0128 	&	 1.37$\pm$0.06\\[0.5ex]
		 8255796 	&	 - 	&	 - 	&	 - 	&	 - 	&	 - 	&	 - 	&	 - 	&	 -\\[0.5ex]
		 8381949 	&	 4.319$\pm$0.005 	&	 3.839$\pm$0.088 	&	 3.708$\pm$0.172 	&	 5.58$\pm$0.91 	&	 0.121$\pm$0.016 	&	 0.220$\pm$0.017 	&	 0.2062$\pm$0.0165 	&	 1.58$\pm$0.12\\[0.5ex]
		 8459899 	&	 4.205$\pm$0.030 	&	 3.859$\pm$0.102 	&	 2.948$\pm$0.057 	&	 3.88$\pm$0.30 	&	 0.092$\pm$0.018 	&	 0.175$\pm$0.053 	&	 0.2006$\pm$0.0967 	&	 2.30$\pm$0.36\\[0.5ex]
		 8714886 	&	 4.247$\pm$0.002 	&	 3.864$\pm$0.009 	&	 3.298$\pm$0.002 	&	 4.78$\pm$0.04 	&	 0.095$\pm$0.001 	&	 0.145$\pm$0.003 	&	 0.0998$\pm$0.0001 	&	 1.68$\pm$0.05\\[0.5ex]
		 8766405 	&	 4.147$\pm$0.009 	&	 3.589$\pm$0.075 	&	 3.034$\pm$0.044 	&	 5.59$\pm$0.27 	&	 0.056$\pm$0.006 	&	 0.124$\pm$0.021 	&	 0.1524$\pm$0.0198 	&	 2.11$\pm$0.19\\[0.5ex]
		 9020774 	&	 4.149$\pm$0.005 	&	 4.152$\pm$0.080 	&	 2.375$\pm$0.098 	&	 2.60$\pm$0.23 	&	 0.124$\pm$0.017 	&	 0.176$\pm$0.026 	&	 0.1101$\pm$0.0436 	&	 2.12$\pm$0.07\\[0.5ex]
		 9715425 	&	 4.202$\pm$0.012 	&	 3.768$\pm$0.063 	&	 3.089$\pm$0.036 	&	 4.63$\pm$0.42 	&	 0.091$\pm$0.011 	&	 0.164$\pm$0.021 	&	 0.1578$\pm$0.0344 	&	 2.13$\pm$0.13\\[0.5ex]
		 10526294 	&	 - 	&	 - 	&	 - 	&	 - 	&	 - 	&	 - 	&	 - 	&	 -\\[0.5ex]
		 10536147 	&	 4.329$\pm$0.009 	&	 4.105$\pm$0.075 	&	 3.405$\pm$0.019 	&	 3.70$\pm$0.17 	&	 0.178$\pm$0.016 	&	 0.274$\pm$0.008 	&	 0.2500$\pm$0.0410 	&	 1.41$\pm$0.57\\[0.5ex]
		 11360704 	&	 4.241$\pm$0.002 	&	 3.900$\pm$0.085 	&	 3.154$\pm$0.065 	&	 4.15$\pm$0.33 	&	 0.119$\pm$0.008 	&	 0.207$\pm$0.033 	&	 0.1842$\pm$0.0744 	&	 1.99$\pm$0.31\\[0.5ex]
		 11971405 	&	 4.192$\pm$0.008 	&	 3.992$\pm$0.002 	&	 2.714$\pm$0.029 	&	 3.14$\pm$0.01 	&	 0.114$\pm$0.004 	&	 0.179$\pm$0.011 	&	 0.1753$\pm$0.0255 	&	 2.38$\pm$0.10\\[0.5ex]
		 12258330 	&	 4.203$\pm$0.002 	&	 4.276$\pm$0.016 	&	 2.485$\pm$0.000 	&	 2.29$\pm$0.02 	&	 0.153$\pm$0.009 	&	 0.220$\pm$0.025 	&	 0.1491$\pm$0.0115 	&	 1.97$\pm$0.10\\[0.5ex]
 \enddata
\end{deluxetable*}
}


\section{Separations of AICc distributions}\label{Sec:Grid_stds}

\noindent In Table~\ref{Tab:stds} we provide a summary of the results displayed in Figs.~\ref{fig:stds_val} and~\ref{fig:grid_stds_hist}, indicating first the $D_{\rm mix} (r)$ profile which \cite{Pedersen2021} estimated to best match the observed period spacing patterns using the $\boldsymbol{\psi}_1, \dots, \boldsymbol{\psi}_8$ notation as summarized in Table~\ref{Tab:Dmix}. The following columns note which of the remaining seven mixing profiles for which the 1000 perturbations of Eq.~(\ref{Eq:stat_mod}) result in AICc distributions whose 86th percentile is separated from the median of the AICc distribution of the best model estimate by less than 1$\sigma_{\rm std}$, or more than one, two, or three $\sigma_{\rm std}$, respectively.

{\movetabledown=1.5in\tabcolsep=10pt
\begin{rotatetable}
\begin{deluxetable*}{cccccc}
\tablecaption{Summary of the capability of the observed period spacing patterns to differentiate between different $D_{\rm mix} (r)$ profiles. For each of the 26 SPB stars the $D_{\rm mix} (r)$ profile of the best model estimate from \cite{Pedersen2021} is first listed, following the notation summarized in Table~\ref{Tab:Dmix}. The third column lists the mixing profiles for which the 86th percentile of their AICc distributions resulting from the 1000 perturbations of Eq.~(\ref{Eq:stat_mod}) are separated from the best matching model by less than $1\sigma_{\rm std}$. The last three columns list the $D_{\rm mix} (r)$ profiles for which the corresponding separation is larger than one, two, and three $\sigma_{\rm std}$, respectively.  \label{Tab:stds}}
\tablewidth{700pt}
\tabletypesize{\small}
\tablehead{
\colhead{KIC} & \colhead{Best model} & 
\colhead{$\sigma_{\rm std} < 1$} & \colhead{$\sigma_{\rm std} \geq 1$} & 
\colhead{$\sigma_{\rm std} \geq 2$} & \colhead{$\sigma_{\rm std} \geq 3$}
} 
\startdata
		 3240411 	&	 $\psi_3$ 	&	  	&	 $\psi_1$, $\psi_2$, $\psi_4$, $\psi_5$, $\psi_6$, $\psi_7$, $\psi_8$ 	&	 $\psi_1$, $\psi_2$, $\psi_4$, $\psi_5$, $\psi_7$, $\psi_8$ 	&	 $\psi_1$, $\psi_2$, $\psi_4$, $\psi_5$, $\psi_7$, $\psi_8$\\[0.5ex]
		 3459297 	&	 $\psi_8$ 	&	 $\psi_3$, $\psi_5$, $\psi_7$ 	&	 $\psi_1$, $\psi_2$, $\psi_4$, $\psi_6$ 	&	 $\psi_1$, $\psi_2$, $\psi_4$, $\psi_6$ 	&	 $\psi_1$, $\psi_2$, $\psi_4$, $\psi_6$\\[0.5ex]
		 3865742 	&	 $\psi_5$ 	&	 $\psi_3$ 	&	 $\psi_1$, $\psi_2$, $\psi_4$, $\psi_6$, $\psi_7$, $\psi_8$ 	&	 $\psi_1$, $\psi_2$, $\psi_4$, $\psi_6$, $\psi_8$ 	&	 $\psi_1$, $\psi_2$, $\psi_4$, $\psi_6$, $\psi_8$\\[0.5ex]
		 4930889 	&	 $\psi_1$ 	&	 $\psi_5$ 	&	 $\psi_2$, $\psi_3$, $\psi_4$, $\psi_6$, $\psi_7$, $\psi_8$ 	&	 $\psi_2$, $\psi_3$, $\psi_6$, $\psi_7$, $\psi_8$ 	&	 $\psi_2$, $\psi_3$, $\psi_6$, $\psi_7$, $\psi_8$\\[0.5ex]
		 4936089 	&	 $\psi_6$ 	&	 $\psi_5$, $\psi_7$ 	&	 $\psi_1$, $\psi_2$, $\psi_3$, $\psi_4$, $\psi_8$ 	&	 $\psi_4$ 	&	 $\psi_4$\\[0.5ex]
		 4939281 	&	 $\psi_7$ 	&	  	&	 $\psi_1$, $\psi_2$, $\psi_3$, $\psi_4$, $\psi_5$, $\psi_6$, $\psi_8$ 	&	 $\psi_1$, $\psi_2$, $\psi_3$, $\psi_4$, $\psi_5$, $\psi_6$, $\psi_8$ 	&	 $\psi_1$, $\psi_2$, $\psi_3$, $\psi_4$, $\psi_5$, $\psi_6$, $\psi_8$\\[0.5ex]
		 5309849 	&	 $\psi_6$ 	&	  	&	 $\psi_1$, $\psi_2$, $\psi_3$, $\psi_4$, $\psi_5$, $\psi_7$, $\psi_8$ 	&	 $\psi_1$, $\psi_2$, $\psi_3$, $\psi_4$, $\psi_5$, $\psi_7$, $\psi_8$ 	&	 $\psi_1$, $\psi_2$, $\psi_3$, $\psi_4$, $\psi_5$, $\psi_7$, $\psi_8$\\[0.5ex]
		 5941844 	&	 $\psi_7$ 	&	 $\psi_1$ 	&	 $\psi_2$, $\psi_3$, $\psi_4$, $\psi_5$, $\psi_6$, $\psi_8$ 	&	 $\psi_2$, $\psi_3$, $\psi_4$, $\psi_5$, $\psi_6$, $\psi_8$ 	&	 $\psi_2$, $\psi_3$, $\psi_5$, $\psi_6$, $\psi_8$\\[0.5ex]
		 6352430 	&	 $\psi_7$ 	&	 $\psi_2$, $\psi_3$, $\psi_4$, $\psi_5$, $\psi_6$, $\psi_8$ 	&	 $\psi_1$ 	&	 $\psi_1$ 	&	 \\[0.5ex]
		 6462033 	&	 $\psi_8$ 	&	 $\psi_7$ 	&	 $\psi_1$, $\psi_2$, $\psi_3$, $\psi_4$, $\psi_5$, $\psi_6$ 	&	 $\psi_1$, $\psi_2$, $\psi_3$, $\psi_4$, $\psi_5$, $\psi_6$ 	&	 $\psi_1$, $\psi_2$, $\psi_3$, $\psi_4$, $\psi_5$\\[0.5ex]
		 6780397 	&	 $\psi_6$ 	&	 $\psi_2$, $\psi_7$ 	&	 $\psi_1$, $\psi_3$, $\psi_4$, $\psi_5$, $\psi_8$ 	&	 $\psi_1$, $\psi_3$, $\psi_4$, $\psi_5$, $\psi_8$ 	&	 $\psi_1$, $\psi_3$, $\psi_4$, $\psi_5$, $\psi_8$\\[0.5ex]
		 7630417 	&	 $\psi_2$ 	&	 $\psi_1$ 	&	 $\psi_3$, $\psi_4$, $\psi_5$, $\psi_6$, $\psi_7$, $\psi_8$ 	&	 $\psi_3$, $\psi_4$, $\psi_5$, $\psi_6$, $\psi_7$, $\psi_8$ 	&	 $\psi_3$, $\psi_4$, $\psi_5$, $\psi_7$, $\psi_8$\\[0.5ex]
		 7760680 	&	 $\psi_7$ 	&	 $\psi_3$ 	&	 $\psi_1$, $\psi_2$, $\psi_4$, $\psi_5$, $\psi_6$, $\psi_8$ 	&	 $\psi_1$, $\psi_2$, $\psi_4$, $\psi_5$, $\psi_6$, $\psi_8$ 	&	 $\psi_1$, $\psi_2$, $\psi_4$, $\psi_5$, $\psi_6$, $\psi_8$\\[0.5ex]
		 8057661 	&	 $\psi_7$ 	&	 $\psi_3$, $\psi_5$, $\psi_6$, $\psi_8$ 	&	 $\psi_1$, $\psi_2$, $\psi_4$ 	&	 $\psi_1$, $\psi_4$ 	&	 $\psi_4$\\[0.5ex]
		 8255796 	&	 $\psi_7$ 	&	  	&	 $\psi_1$, $\psi_2$, $\psi_3$, $\psi_4$, $\psi_5$, $\psi_6$, $\psi_8$ 	&	 $\psi_1$, $\psi_2$, $\psi_3$, $\psi_4$, $\psi_5$, $\psi_6$, $\psi_8$ 	&	 $\psi_1$, $\psi_3$, $\psi_4$, $\psi_5$, $\psi_6$, $\psi_8$\\[0.5ex]
		 8381949 	&	 $\psi_8$ 	&	 $\psi_1$, $\psi_3$ 	&	 $\psi_2$, $\psi_4$, $\psi_5$, $\psi_6$, $\psi_7$ 	&	 $\psi_2$, $\psi_4$, $\psi_5$, $\psi_6$, $\psi_7$ 	&	 $\psi_2$, $\psi_4$, $\psi_5$, $\psi_6$, $\psi_7$\\[0.5ex]
		 8459899 	&	 $\psi_7$ 	&	 $\psi_3$, $\psi_8$ 	&	 $\psi_1$, $\psi_2$, $\psi_4$, $\psi_5$, $\psi_6$ 	&	 $\psi_1$, $\psi_2$, $\psi_4$, $\psi_6$ 	&	 $\psi_1$, $\psi_2$, $\psi_4$, $\psi_6$\\[0.5ex]
		 8714886 	&	 $\psi_6$ 	&	 $\psi_5$ 	&	 $\psi_1$, $\psi_2$, $\psi_3$, $\psi_4$, $\psi_7$, $\psi_8$ 	&	 $\psi_1$, $\psi_2$, $\psi_3$, $\psi_4$, $\psi_8$ 	&	 $\psi_1$, $\psi_2$, $\psi_3$, $\psi_4$, $\psi_8$\\[0.5ex]
		 8766405 	&	 $\psi_7$ 	&	 $\psi_2$, $\psi_4$, $\psi_6$, $\psi_8$ 	&	 $\psi_1$, $\psi_3$, $\psi_5$ 	&	 $\psi_1$, $\psi_3$ 	&	 $\psi_1$\\[0.5ex]
		 9020774 	&	 $\psi_3$ 	&	 $\psi_1$, $\psi_8$ 	&	 $\psi_2$, $\psi_4$, $\psi_5$, $\psi_6$, $\psi_7$ 	&	 $\psi_2$, $\psi_4$, $\psi_6$, $\psi_7$ 	&	 $\psi_2$, $\psi_4$, $\psi_6$, $\psi_7$\\[0.5ex]
		 9715425 	&	 $\psi_1$ 	&	 $\psi_7$ 	&	 $\psi_2$, $\psi_3$, $\psi_4$, $\psi_5$, $\psi_6$, $\psi_8$ 	&	 $\psi_5$, $\psi_6$, $\psi_8$ 	&	 $\psi_6$\\[0.5ex]
		 10526294 	&	 $\psi_1$ 	&	  	&	 $\psi_2$, $\psi_3$, $\psi_4$, $\psi_5$, $\psi_6$, $\psi_7$, $\psi_8$ 	&	 $\psi_2$, $\psi_3$, $\psi_4$, $\psi_5$, $\psi_6$, $\psi_7$, $\psi_8$ 	&	 $\psi_2$, $\psi_3$, $\psi_4$, $\psi_5$, $\psi_6$, $\psi_7$, $\psi_8$\\[0.5ex]
		 10536147 	&	 $\psi_3$ 	&	 $\psi_1$, $\psi_2$, $\psi_4$, $\psi_5$ 	&	 $\psi_6$, $\psi_7$, $\psi_8$ 	&	 $\psi_6$, $\psi_8$ 	&	 $\psi_6$, $\psi_8$\\[0.5ex]
		 11360704 	&	 $\psi_4$ 	&	 $\psi_1$, $\psi_2$, $\psi_5$, $\psi_6$, $\psi_7$, $\psi_8$ 	&	 $\psi_3$ 	&	  	&	 \\[0.5ex]
		 11971405 	&	 $\psi_5$ 	&	 $\psi_2$ 	&	 $\psi_1$, $\psi_3$, $\psi_4$, $\psi_6$, $\psi_7$, $\psi_8$ 	&	 $\psi_1$, $\psi_3$, $\psi_4$, $\psi_6$, $\psi_7$, $\psi_8$ 	&	 $\psi_1$, $\psi_3$, $\psi_4$, $\psi_6$, $\psi_7$, $\psi_8$\\[0.5ex]
		 12258330 	&	 $\psi_2$ 	&	 $\psi_3$ 	&	 $\psi_1$, $\psi_4$, $\psi_5$, $\psi_6$, $\psi_7$, $\psi_8$ 	&	 $\psi_1$, $\psi_5$, $\psi_6$, $\psi_7$, $\psi_8$ 	&	 $\psi_5$, $\psi_6$, $\psi_7$, $\psi_8$\\[0.5ex]
 \enddata
\end{deluxetable*}
\end{rotatetable}
}

\clearpage
\section{Grid vs statistical model comparison}\label{Sec:Grid_vs_stat}

\begin{figure*}
\includegraphics[width=\linewidth]{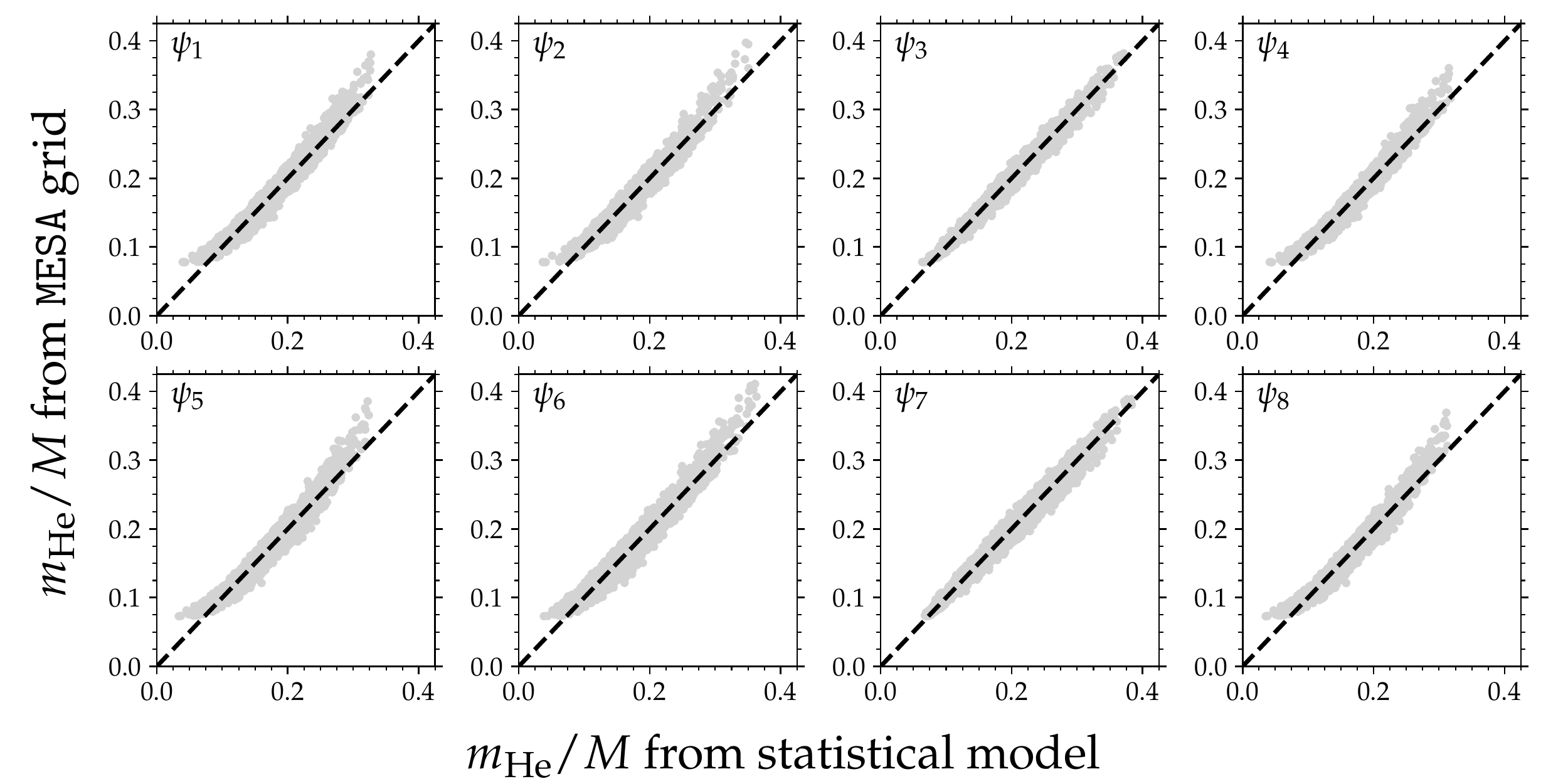}
\caption{Comparison of helium core masses obtained from the \texttt{MESA} computations for the eight different grids (x-axis) and those predicted by the statistical models (y-axis). The black dashed lines indicate where the two are equal. Different subplots correspond to a different grid, i.e. a different combination of the convective core boundary mixing and envelope mixing as indicated in the upper left corner of each subplot. }
	\label{fig:Grid_vs_stat}
\end{figure*}

\noindent In Sect.~\ref{sec:Hecore} we discussed how the helium core masses obtained at the end of the main-sequence evolution for the 26 SPB stars are estimated using statistical models. Figure~\ref{fig:Grid_vs_stat} provides a comparison between the $m_{\rm He}$ values derived with \texttt{MESA} to those estimated from the statistical models for the original varied $\boldsymbol{\theta}$ parameters in the eight $\boldsymbol{\psi}_1, \dots, \boldsymbol{\psi}_8$ grids.
For both grid $\psi_3$ and $\psi_7$ with vertical shear mixing in the radiative envelope the statistical models do better at reproducing the \texttt{MESA} grid values at both the lower and higher mass end of the helium cores. This is because more terms were included in the statistical models for these two grids than for the remaining six grids, where the Bayesian Information Criterion (BIC) values started to increase when additional terms were added to the statistical model. The BIC values were used to determine the number of fractional polynomial terms which should be included in the statistical models, as explained in detail by \cite{Pedersen2021}.

\begin{figure*}
\includegraphics[width=\linewidth]{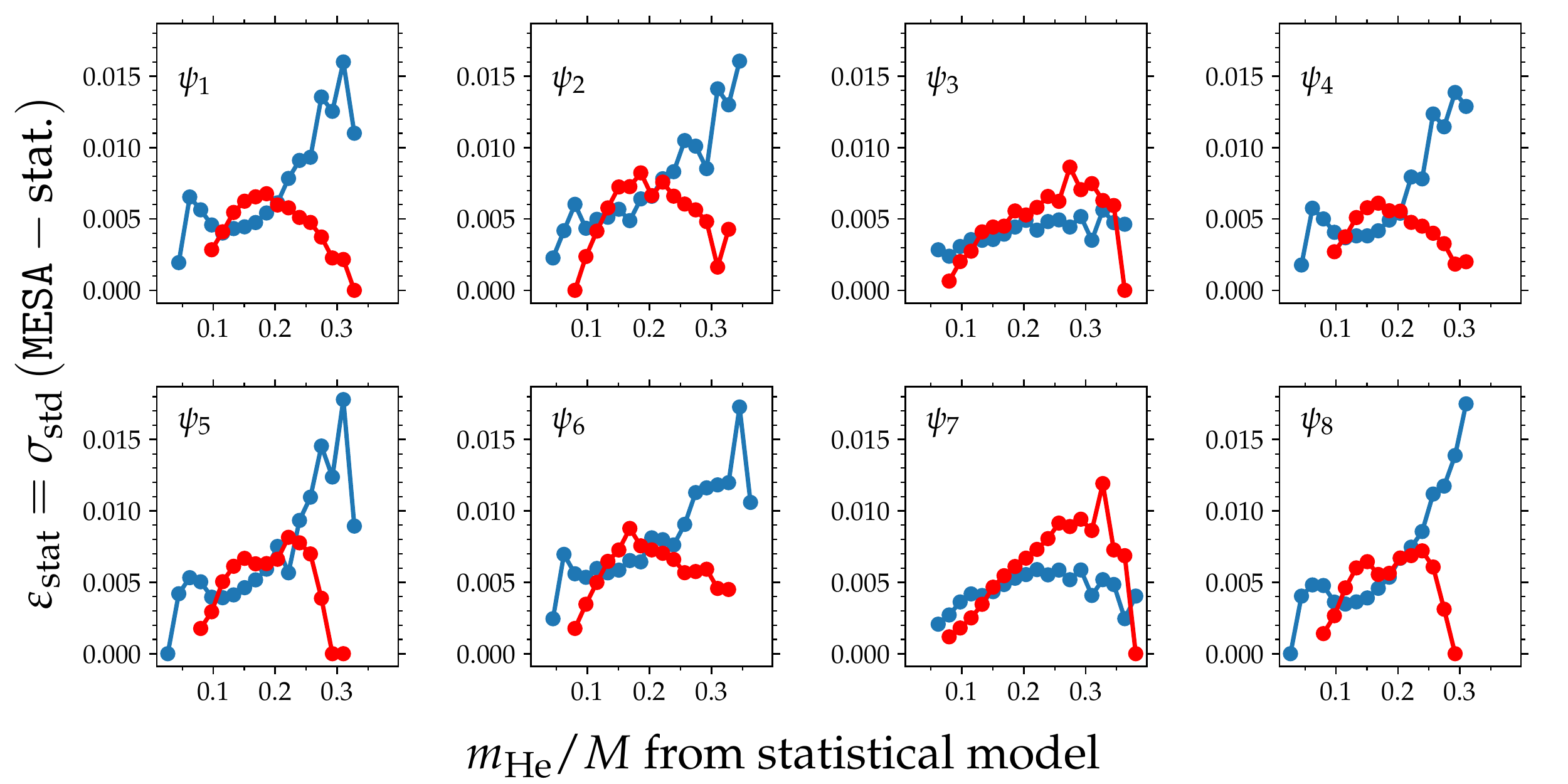}
\caption{Standard deviations of the $\left| m_{\rm He}^{\rm Stat} - m_{\rm He}^\texttt{MESA} \right|$ values derived from the data points in Fig.~\ref{fig:Grid_vs_stat} as a function of $m_{\rm He}$. The standard deviations were derived separately for $m_{\rm He}^{\rm Stat} \geq m_{\rm He}^\texttt{MESA}$ (blue) and $m_{\rm He}^{\rm Stat} < m_{\rm He}^\texttt{MESA}$ (red). Hence the blue curves correspond to upper errors on $m_{\rm He}$ while the red curves are the lower errors resulting from using a statistical model to estimate $m_{\rm He}$.}
	\label{fig:stat_extra_error}
\end{figure*}

As discussed in Appendix~\ref{Sec:av_std_parameters}, the spread in the $m_{\rm He}$ values around the black dashed $m_{\rm He}^{\rm Stat} = m_{\rm He}^\texttt{MESA}$ lines in Fig.~\ref{fig:Grid_vs_stat} were used to estimate and include an additional statistical error term $\varepsilon^{\rm stat}$ on the $m_{\rm He}$ values derived from the statistical models. As the spread in $m_{\rm He}$ around the black dashed lines is not uniform and thereby $m_{\rm He}$ dependent, the standard deviations of the points above and below this line was calculated separately and used as estimates of the upper and lower errors on the estimated helium core masses. In practice, this was done by calculating the standard deviations of $\left| m_{\rm He}^{\rm Stat} - m_{\rm He}^\texttt{MESA} \right|$ separately for $m_{\rm He}^{\rm Stat} \geq m_{\rm He}^\texttt{MESA}$ and $m_{\rm He}^{\rm Stat} < m_{\rm He}^\texttt{MESA}$, corresponding to the blue and red curves in Fig.~\ref{fig:stat_extra_error}, respectively. The final upper and lower $\varepsilon^{\rm stat}$ errors are then obtained by interpolating the estimated $m_{\rm He}$ values onto the blue and red curves in Fig.~\ref{fig:stat_extra_error}.

\clearpage
\section{Definition of helium core mass}\label{Sec:HeCoreDef}

\begin{figure}
\begin{center}
\includegraphics[width=\linewidth]{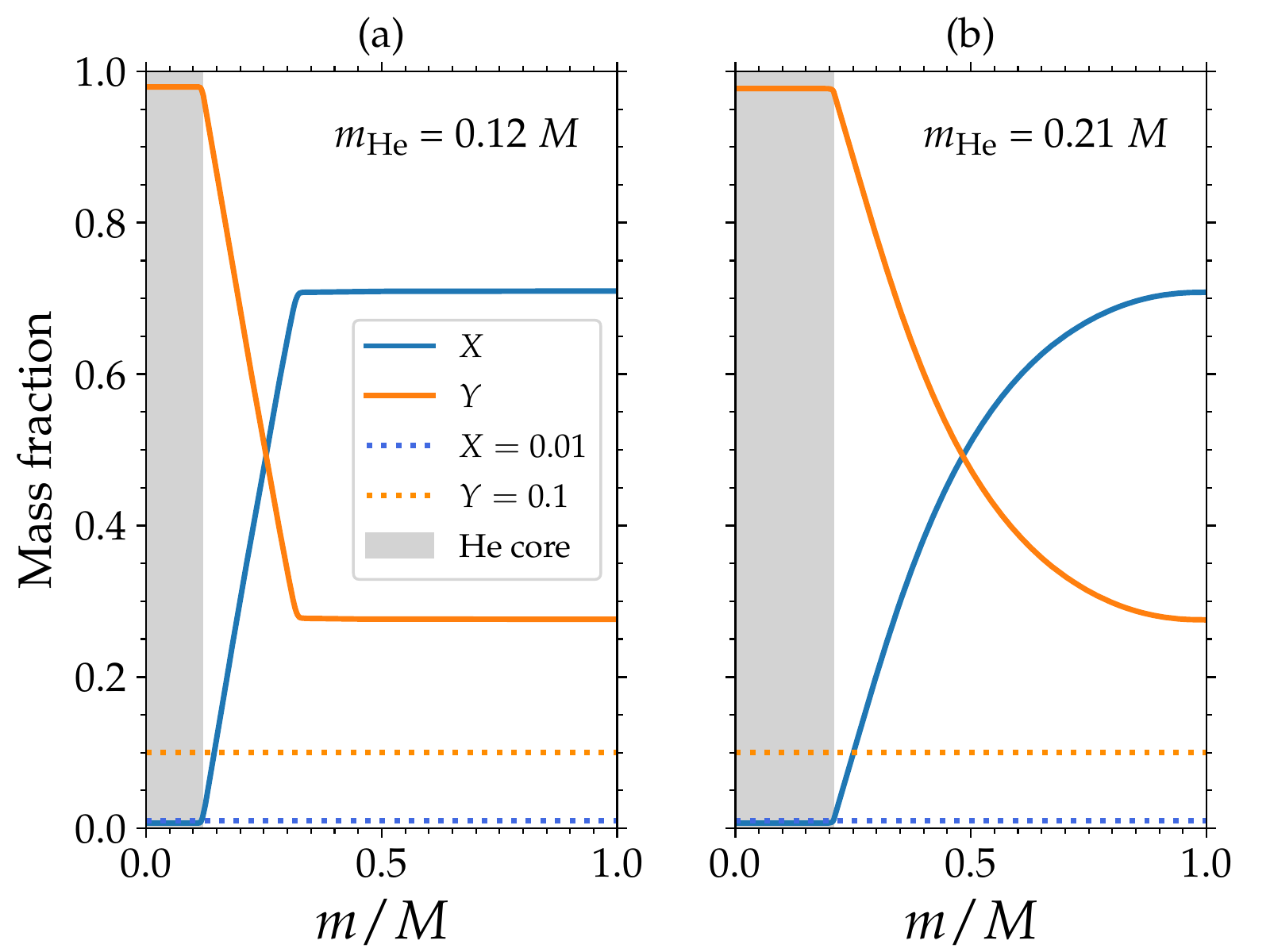}
\caption{Illustration of how the helium core masses are defined by default in \texttt{MESA} according to the hydrogen ($X$, blue) and helium ($Y$, orange) mass fractions for two different stellar models at $X_{\rm c}/X_{\rm ini} = 0.01$. The grey shaded region shows the size of the helium core. The model in panel (a) has the input parameters:  $M = 5,23$\,M$_\odot$, $Z = 0.0142$, $f_{\rm ov} = 0.014$, and $\log D_{\rm env,0} = 1.49$\,cm$^2$\,s$^{-1}$. For the model in panel (b) the input parameters are: $M = 5.2$\,M$_\odot$, $Z = 0.0165$, $f_{\rm ov} = 0.034$, and $\log D_{\rm env,0} = 5.04$\,cm$^2$\,s$^{-1}$. The helium core mass is defined to be the outermost mass coordinate $m/M$ which fulfills both $X < 0.01$ and $Y > 0.1$.}\label{fig:HeCore_illustration}
\end{center}
\end{figure}

\noindent The helium core masses discussed in this work are the ones obtained using the default setup in \texttt{MESA} version r12115. The definition is illustrated in Fig.~\ref{fig:HeCore_illustration} for two different stellar models at the end of the main-sequence evolution ($X_{\rm c}/X_{\rm ini} = 0.01$), which are similar in mass but have experienced vastly different internal mixing throughout their main-sequence lifetimes. The helium core mass is defined according to the hydrogen $X$ and helium $Y$ mass fractions shown in blue and orange in Fig.~\ref{fig:HeCore_illustration}, respectively, and is the outermost mass coordinate $m/M$ where both $X < 0.01$ and $Y > 0.1$ are fulfilled. The helium core corresponds to the inner grey shaded region in the figure, and the mass of the helium core is listed in each subplot of the figure. For both models an exponentially decaying diffusive convective core boundary mixing combined with constant envelope mixing ($\psi_1$) was assumed. As can be seen in the figure, the helium core mass is twice as large in the case of the higher mixing parameters, once again emphasizing the need of taking mixing into account when estimating these masses.

\end{document}